\documentclass[a4paper,fleqn,usenatbib,useAMS]{mnras}
\usepackage{txfonts}
\usepackage[final]{showlabels}
\usepackage{graphicx}
\usepackage{subfig}
\usepackage{threeparttable}

\title[Luminous AGB variables in NGC~3109]{Luminous AGB variables in the dwarf Irregular Galaxy, NGC~3109}

\author[Menzies et al.]{John W. Menzies$^{1}$, Patricia A. Whitelock$^{1,2}$,
Michael W. Feast$^{2,1}$ and Noriyuki Matsunaga$^3$\\
      $^1$ South African Astronomical Observatory, P.O.Box 9, 7935
		                  Observatory, South Africa.\\
      $^2$ Astronomy Department,
                       University of Cape Town, 7701 Rondebosch, South Africa.\\
      $^3$  department of Astronomy, The University of Tokyo, 7-3-1 Hongo, Bunkyo-ku, Tokyo 113-0033, Japan.
						    }

\begin{document}
\maketitle
\begin{abstract}
	In a shallow near-infrared survey of the dwarf Irregular galaxy, NGC~3109, near the periphery of the Local Group,  we have found eight Mira variables, seven of which appear to be oxygen-rich (O-Miras). The periods range from about 430 days to almost 1500 days. Because of our relatively bright limiting magnitude, only 45 of the more than 400 known carbon stars were measured, but none was found to be a large amplitude variable. One of the Miras may be an unrecognised C star. Five of the O-Miras are probably hot-bottom burning stars considering that they are brighter than expected from the period--luminosity relation of Miras and that, by comparison with theoretical evolutionary tracks, they appear to have masses $\ga 4 M_\odot$. 
	A census of very long period ($P>1000$ days) Miras in the Galaxy and Magellanic Clouds is presented and discussed together with the newly discovered long period, but relatively blue, variables in NGC~3109. New $JHKL$ photometry is presented for three O-rich long period Miras in the SMC (including a candidate 
super-AGB star).
\end{abstract}
\section{Introduction}
NGC~3109 is a small galaxy, classified as SB(s)m \citep{deVaucouleurs_1991} or dIrr \citep{McConnachie_2012} that is associated with a group of dwarf irregular galaxies including Sextans A, Sextans B, Antlia and possibly Leo P.  NGC~3109 is outside the virial radius of the Local Group and is probably not bound to it, though it may have been near the Milky Way about 7 -- 9 Gyr ago \citep{Pawlowski_2014}. It was nevertheless included in our infrared search at SAAO for long period variables in the Local Group \citep[see e.g.][]{Menzies_2015}.

Based on the colour of the red giant branch (RGB) \citep{Hidalgo_2008}, the metallicity is very low, $\rm[Fe/H] = -1.84\pm0.2$. Nevertheless, a spectroscopic study of blue supergiants and H\textsc{ii} regions in NGC~3109 \citep{Evans_2007,Hosek_2014,Tramper_2014} results in a metallicity that is significantly higher, $\rm[Fe/H] = -0.67$. \citet{Pena_2007} find $\rm O/H + 12 = 7.77\pm0.07$ for H\textsc{ii} regions, in good agreement with that found by \citet{Evans_2007}, and suggest that the current interstellar medium in NGC~3109 is chemically very homogeneous. Thus the young component of NGC~3109 has an O/H ratio about half that of the SMC.

There have been many determinations of distance, which are summarised by \citet{Soszynski_2006}. We use the modulus, $\rm(m-M)_0=25.6$, in this paper. The reddening is small, $\rm E_{B-V} \sim 0.09$, with about half of this being ascribed to position-dependent internal reddening \citep{Soszynski_2006}. We ignore reddening in the near infrared in the following discussion.

 The galaxy is gas-rich, with an extensive H\textsc{i} envelope \citep{Jobin_1990, Carignan_2013}. The H\textsc{i} is warped in the outer disk towards the neighbouring Antlia galaxy to the south-west suggesting an interaction in the past, though not at present \citep{Carignan_2013}; the disk is seen almost edge-on with an inclination of 75\degr. The rotation curve is consistent with a dark-matter halo that dominates at all radii.
\citet{Demers_2003} found more than 400 C stars in the galaxy, mostly confined to the plane of the disk, and there does not appear to be an extensive halo. They find a C/M ratio of 1.75, typical for a metal-poor galaxy such as this \citep[see e.g.][]{Cioni_2009}.

The present investigation extends our study of AGB variables in the Local Group, which
to date has covered dwarf spheroidals (Leo I: \citep{Menzies_2010}, Fornax: \citep{Whitelock_2009}, Sculptor: \citep{Menzies_2011}) and dwarf
irregulars (NGC 6822: \citep{Whitelock_2013}, IC 1613: \citep{Menzies_2015}, Sgr DIG: \citep{Whitelock_2018}). Our project is to investigate AGB variables
in different environments with two primary aims: first, to test models
of AGB  evolution \citep[e.g.][]{Marigo_2017},  and secondly to investigate how
well Miras can be used as fundamental distance indicators. Short period
Miras, which occur in Galactic globular clusters \citep[e.g.][]{Feast_2002} are
reasonably well understood and are proving to be useful distance
indicators \citep[e.g.][]{Whitelock_2008, Whitelock_2014, Huang_2018,
Yuan_2018}, that will be particularly important in the JWST era,
because they are bright infrared sources.  

Miras with periods
significantly over 400 days, which have more massive progenitors, remain
very poorly understood. These are often strong mid-infrared sources
\citep[][and references therein]{Riebel_2015, Whitelock_2017, Boyer_2017} and are
important in terms of the global dust budget for galaxies. Some of these
undergo hot bottom burning (HBB) \citep{Whitelock_2003}, but how this
depends on their environment remains poorly defined. Of particular current interest is the upper mass limit for AGB stars,
understood to be in the range $6.5-11\ M_{\odot}$ \citep{Karakas_2017},
that determines which stars become supernovae and which end their lives
as white dwarfs.  The input physics is uncertain, so models differ
significantly from one to another. Observations have yet to isolate a
definite example, but the best candidates are the very luminous, large
amplitude, long-period variables of the type we investigate here.

Given the distance to NGC~3109, we are only sensitive to the brightest variables, so our objective in the present study is to investigate the most massive of the AGB variables. A deeper survey is required to get a complete sample. 

\section{Observations}
Our observations were made with the SIRIUS camera on the Japanese-SAAO IRSF telescope at Sutherland. The camera produces 
simultaneous $J, H$ and $K_S$ images covering a 7.2$\times$7.2 arcmin square field with a scale of 0.45 arcsec/pixel. NGC~3109 is extended in approximately the east-west direction and it was necessary to use 3 pointings to cover the bulk of the galaxy. These were centred at $\alpha$(2000.0)=10:03:06.8,\,$\delta$(2000.0)=--26:09:31.9, (field 1), and displaced by $\pm7.2$ arcmin (fields 2 and 3, respectively) in RA.
The aim of this observational series 
was to find long-period variables; observations were made at 17 epochs spread over a little more than 3 years.
For each of the 3 fields, 10 dithered images were combined after flatfielding and dark and sky subtraction. Typical exposures were 
of either 20 or 30 seconds' duration, depending on the seeing and on the brightness of the sky in the $K_S$ band, and typically 50 or 30 exposures, respectively, were combined to produce the final images. Photometry was
performed using \textsc{dophot} in `fixed-position' mode, using the best-seeing $H$-band image as a template. Aladin \citep{Bonnarel_2000}
was used to correct the WCS on each template and RA and Dec were determined for each measured star. This allowed a
cross-correlation to be made with the 2MASS catalogue, and photometric zero points were determined by comparison of our photometry with that of 2MASS. With at least 24, mostly foreground, stars in the range $K_S=11.7 - 14.5$ mag, in common in each field, the standard errors of the zero points in the three filters are $\pm0.01$ mag.
\section{Colour-Magnitude and Colour-Colour Diagrams}

\begin{figure*}
\begin{tabular}{cc}
 \includegraphics[width=0.50\textwidth]{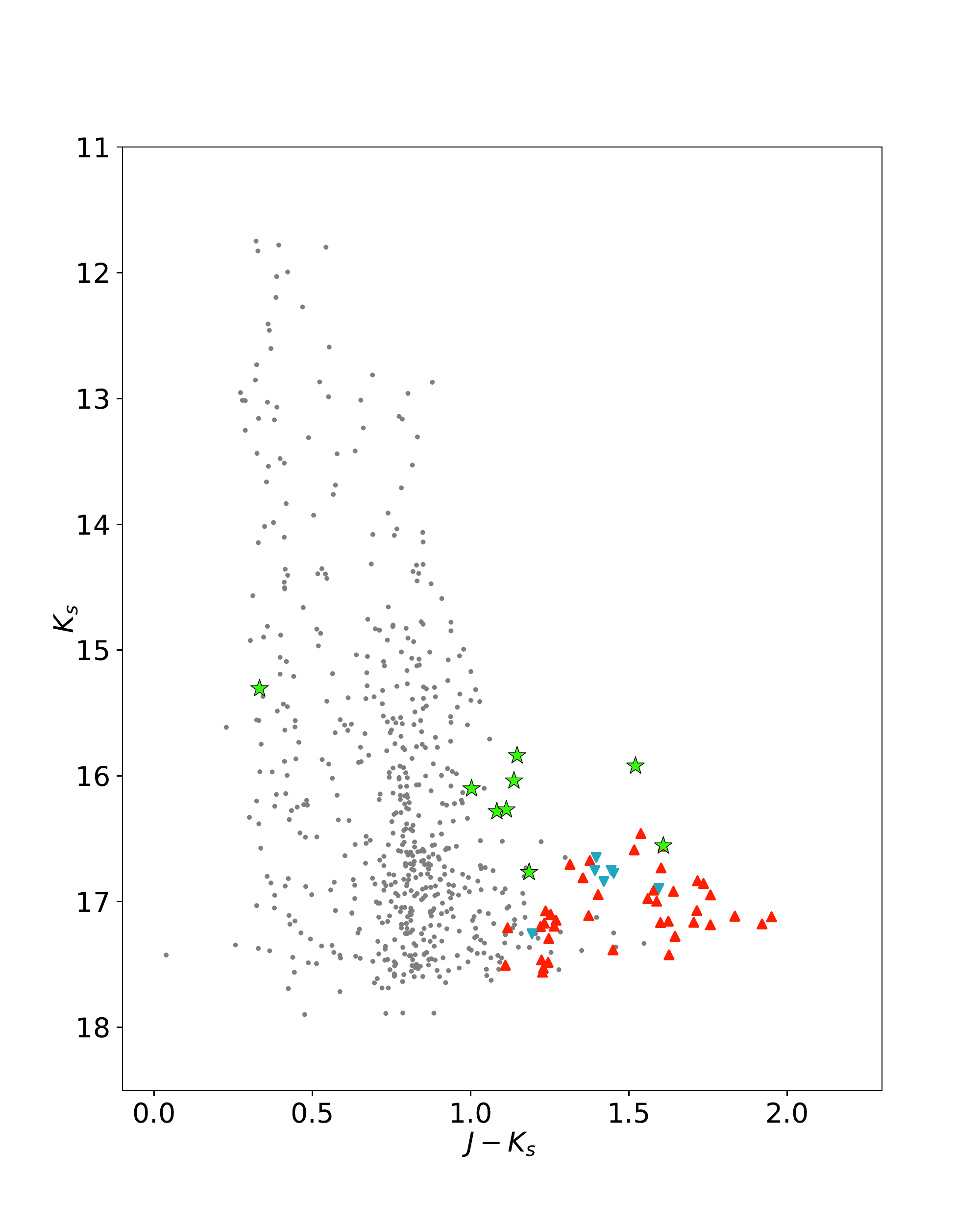} &   \includegraphics[width=0.50\textwidth]{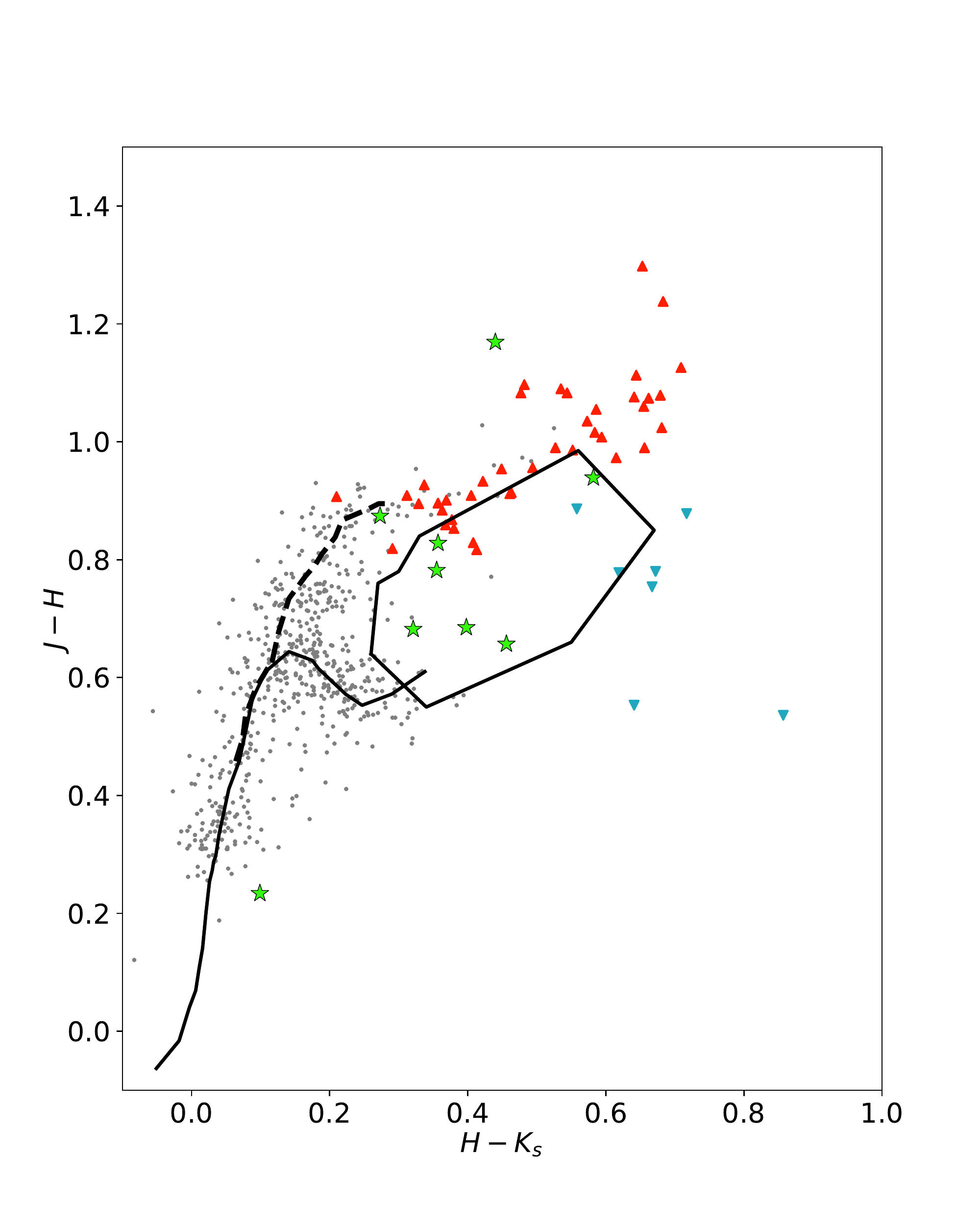} \\
 \end{tabular}
	\caption{(Left) Colour-magnitude diagram for NGC~3109. The small dots (grey) show the field and galaxy non-variable stars. Triangles (red) show known or suspected C stars, star symbols (green) are the variables listed in Table~\ref{tab_N3109var}. On the basis of their positions in the two-colour diagram the points marked as inverted triangles (cyan) are probably background galaxies. (Right) Two-colour diagram for NGC~3109. Coloured symbols are as in the colour-magnitude diagram. The continuous and dashed lines show the loci of field dwarfs and giants, respectively, from \citet{Bess_1988} transformed from the Glass system to the 2MASS system. The region occupied by O-rich Miras in the Galaxy is shown by the closed box \citep{Feast_1984}, the so-called `Mira coffin'.}
\label{fig_SAAOphot}
\end{figure*}

Fig.~\ref{fig_SAAOphot} shows the $K_S-(J-K_S)$ and  $(J-H)-(H-K_S)$ diagrams. The variables and C stars are highlighted. In the two-colour diagram, the box outline, or `Mira coffin' \citep{Feast_1984}, shows the region where O-rich Miras in the Galaxy are found (the original coordinates of the box edges were on the SAAO system and have been transformed to the 2MASS system for this diagram using relations in \citet{Carpenter_2001}). The box is displaced with respect to the locus of normal giants in both $J-H$ and $H-K_S$ due to the dominance of H$_2$O absorption in the near infrared. Galactic carbon-rich stars (including Miras), with little or no water absorption, fall along the upper left edge of the box, and extend it to larger $J-H$ and $H-K_S$ when they experience significant circumstellar reddening 
\citep{Feast1982}.
Six of the newly-discovered Miras fall in the box.

Of the more than 400 C stars found in the galaxy by \citet{Demers_2003} only 45 appear in our photometric catalogue because of our relatively bright limiting magnitude. A colour-magnitude diagram published by \citet{Gorski_2011} shows that the bulk of the C stars must be about one magnitude fainter than our observing limit.

\subsection{Separation of Field and Member stars}
\begin{figure*}
\begin{tabular}{cc}
 \includegraphics[width=0.50\textwidth]{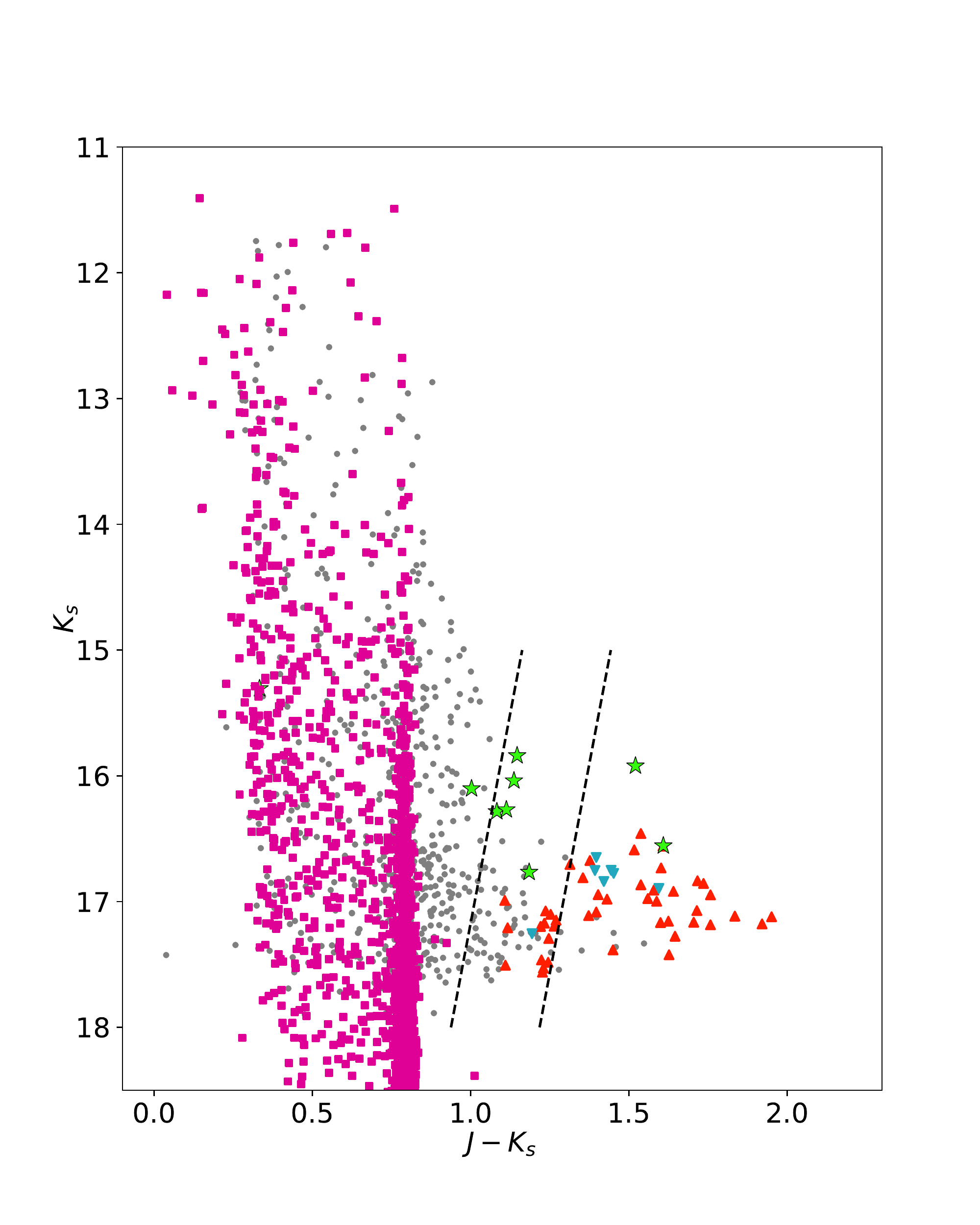} &   \includegraphics[width=0.50\textwidth]{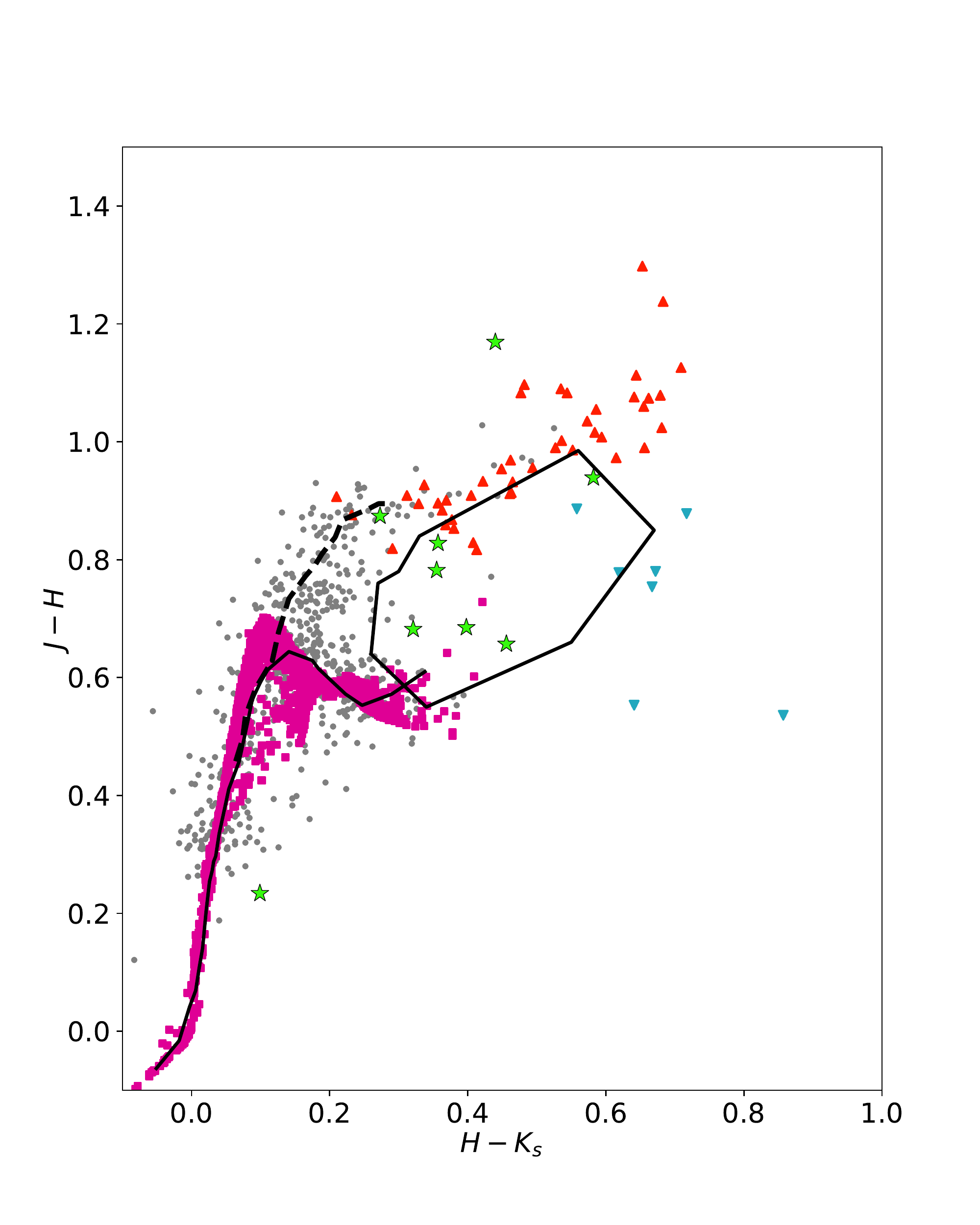} \\
 \end{tabular}
	\caption{Colour-magnitude and two-colour diagrams for NGC~3109 with stars from a \textsc{trilegal} model superimposed as squares (magenta). Symbols and colours otherwise as in Fig.~\ref{fig_SAAOphot}. The region between the dashed lines is where O-rich stars on the AGB with $\rm[Fe/H] = -1.2$ should be found according to \citet{Cioni_2006} (see section~\ref{sec:Cstars} for discussion).} 
\label{fig_SAAOphot_Tri}
\end{figure*}

From the distribution of stars in the two diagrams, it seems clear that there are many field stars present. It is instructive to compare these diagrams with what is expected in the field based on a \textsc{trilegal} \citep{Girardi_2005} simulation for the same area as our photometry covers. This is shown in Fig.~\ref{fig_SAAOphot_Tri}, where the \textsc{trilegal} points are shown in magenta superposed on our diagrams. It should be noted that no observational errors have been applied to the \textsc{trilegal} data.

It appears that, in our sample of the NGC~3109 field, unambiguous members lie to the red of $J-K_S\sim0.9$. Thus, as expected, our survey is very shallow and comprises mostly the late-type giants on the AGB and supergiants (if present). The same conclusion could be reached by making $J-H$ and $H-K_S$ cuts in the two-colour diagram to isolate the obvious field dwarf and giant sequences.

\section{Variable Stars}

As stated earlier, the purpose of our study was to search for and characterise long period variables in NGC~3109. We have discovered nine large amplitude variables -- eight are apparently Miras, while one is unusual, being blue and having a peculiar light curve as discussed
in section~\ref{ssec:3}.

\subsection{Variables found in our catalogue}\label{ssec:1}

\begin{table*}
\caption{$JHK_S$ Photometry of Nine Large Amplitude Variables}
\begin{threeparttable}
\centering
\begin{tabular}{cccccccccccccc}

Name & RA & Dec & Period & $\Delta J$ & $\Delta H$ & $\Delta K_S$ & $K_S$ & $J-H$ & $H-K_S$ & $J-K_S$ & BC& m$_{bol}$ & note \\
	& \multicolumn{2}{c}{(2000.0)}&  day   & \multicolumn{9}{c}{\xleaders\hbox{--}\hfill\kern0pt mag\xleaders\hbox{--}\hfill\kern0pt}\\
1067 & 150.73837 & -26.14197 &  591   & 0.88 & 1.22 & 1.00 & 16.102 & 0.682 & 0.321 & 1.003 &3.20   &19.30  & \\
1112 & 150.80582 & -26.15483 &  583 & 0.49 & 0.59 & 0.62 & 16.039 & 0.782 & 0.355 & 1.137 &3.20   &19.24  & \\
1153 & 150.75345 & -26.17370 &  678   & 0.91 & 0.97 & 0.93 & 16.284 & 0.685 & 0.398 & 1.083 &3.20   &19.48  & \\
1224 & 150.82961 & -26.14808 &  680 & 0.73 & 0.85 & 0.74 & 16.269 & 0.657 & 0.456 & 1.113 &3.20   &19.47  & \\
2075 & 150.91414 & -26.17125 & 1109 & 1.69 & 1.34 & 1.34 & 16.556 & 1.169 & 0.440 & 1.609 &3.19   &19.74  & \\
2081 & 150.86487 & -26.15757 &  434 & 0.71 & 0.74 & 0.60 & 16.766 & 0.828 & 0.357 & 1.185 &3.20   &19.97  & \\
3064 & 150.65652 & -26.14415 &  562 & 0.37 & 0.42 & 0.47 & 15.839 & 0.874 & 0.273 & 1.147 &3.20   &19.04  & \\
3089 & 150.59729 & -26.17039 & 1486 & 1.24 & 1.18 & 0.98 & 15.921 & 0.939 & 0.582 & 1.521 &3.19   &19.11  &\tnote{a} \\
1060 & 150.75887 & -26.14944 &  276   & 0.08 & 0.10 & 0.03 & 15.306 & 0.234 & 0.099 & 0.333 & --   &--  &\tnote{b}\\
\end{tabular}
       \begin{tablenotes}
            \item[a] period about the same as the length of the data train.
	    \item[b] Period and amplitudes after removal of linear trend. See text in section~\ref{ssec:3}.
        \end{tablenotes}
\end{threeparttable}
\label{tab_N3109var}
\end{table*}

\begin{figure}
	\includegraphics[width=0.47\textwidth]{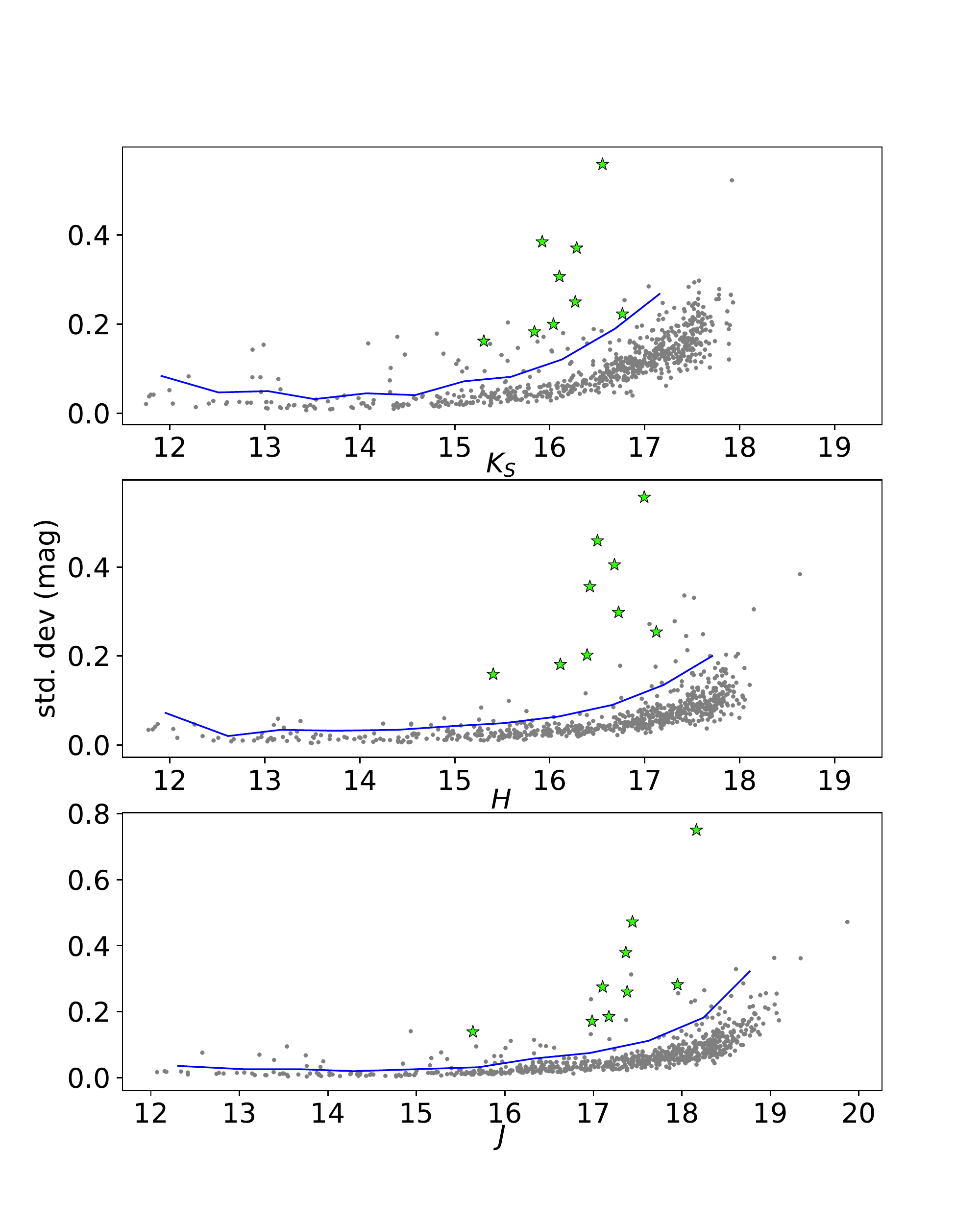}
	\caption{Standard deviation versus magnitude for the catalogue. Periodic variables and star \#1060 are marked as star symbols (green). Variability was tested for stars with standard deviations above the lines simultaneously in all wavebands; see section~\ref{ssec:1}.}
	\label{fig_stddev}
\end{figure}

Median standard deviations were determined at half-magnitude intervals for each of the wavebands represented in Fig.~\ref{fig_stddev}. Lines were drawn at twice these median values (see the figure) and a star was considered variable if for all of $J$, $H$ and $K_S$ its standard deviation lay above the relevant line.
We used Lomb-Scargle periodograms to find possible periods. Because the data points are not well distributed in time and for some objects are relatively noisy, we found that the best
results were obtained by adopting the multiband periodogram approach of \citet{VanderPlas_2015} whereby all 3 wavebands are fitted simultaneously. 
The stars for which we found periods are shown as asterisks in Fig.~\ref{fig_stddev}, where we plot standard deviation against magnitude for our catalogue. The stars with similarly large standard deviations may be variable but are either not periodic or may actually be marginally resolved background galaxies that appear to vary because of seeing fluctuations from image to image.

\begin{figure*}
\begin{tabular}{cc}
 \includegraphics[width=0.42\textwidth]{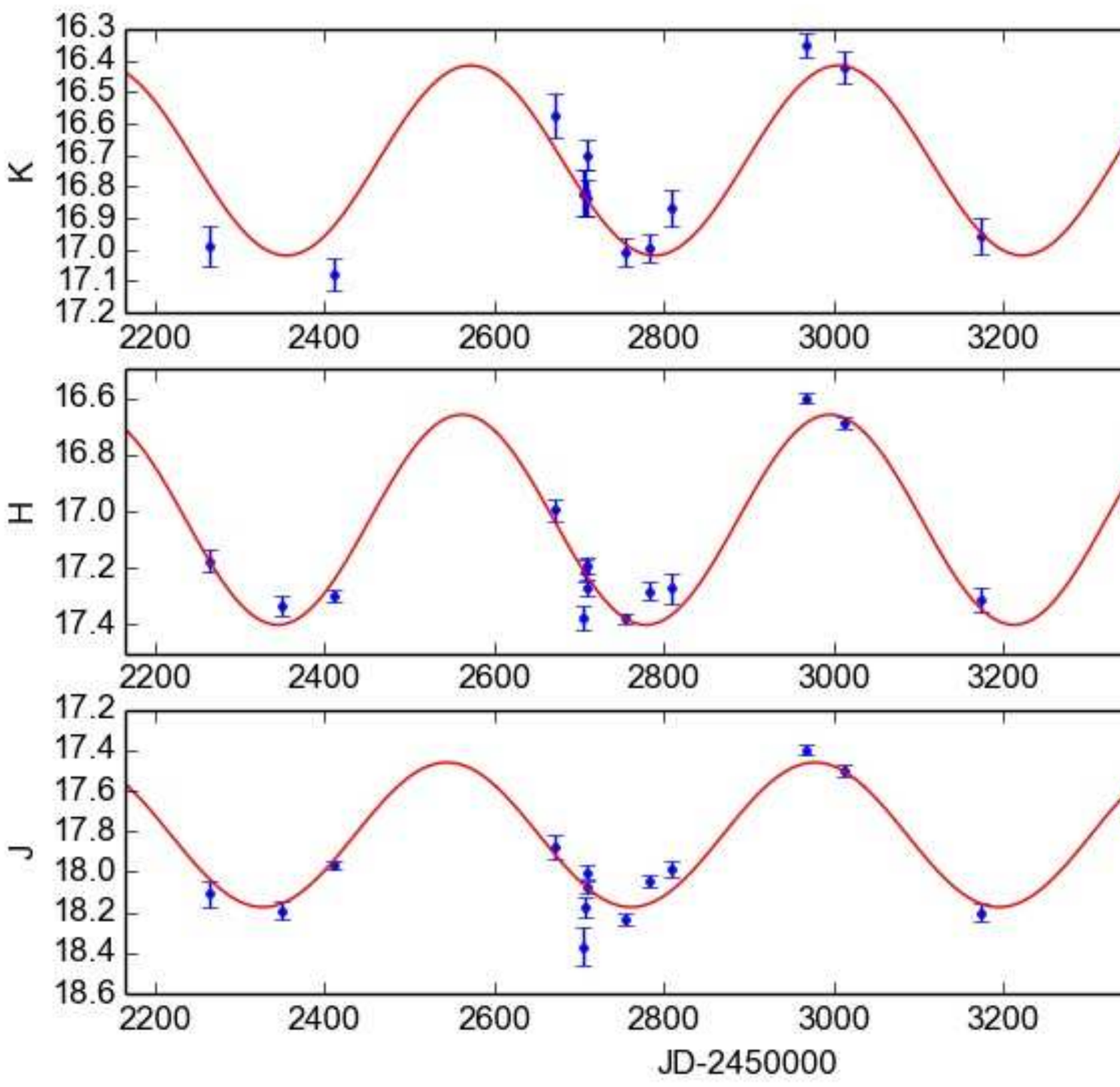} &   \includegraphics[width=0.42\textwidth]{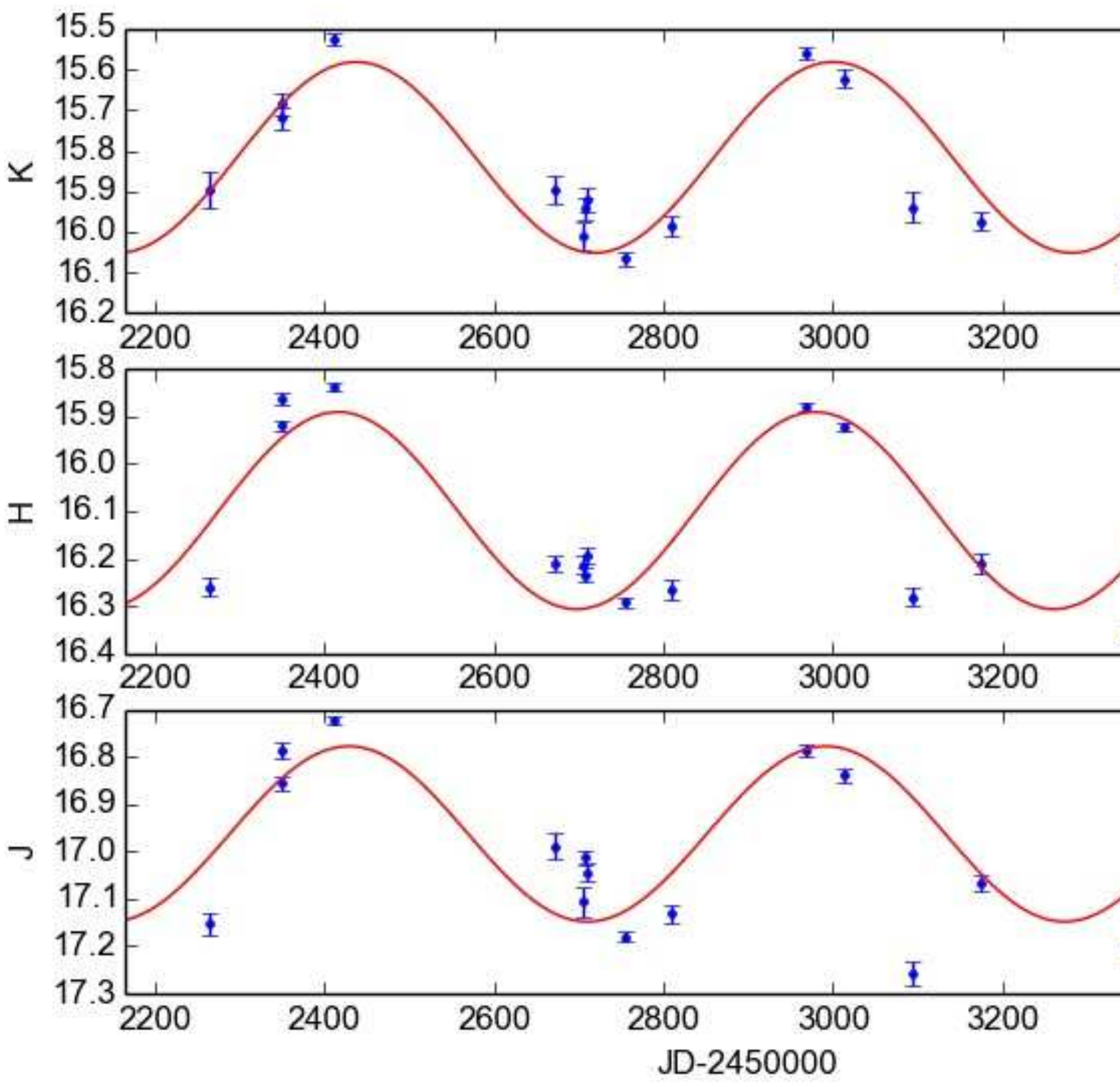} \\
	(a) 2081 (434 d)  & (b) 3064  (562 d) \\
[-2pt]
 \includegraphics[width=0.42\textwidth]{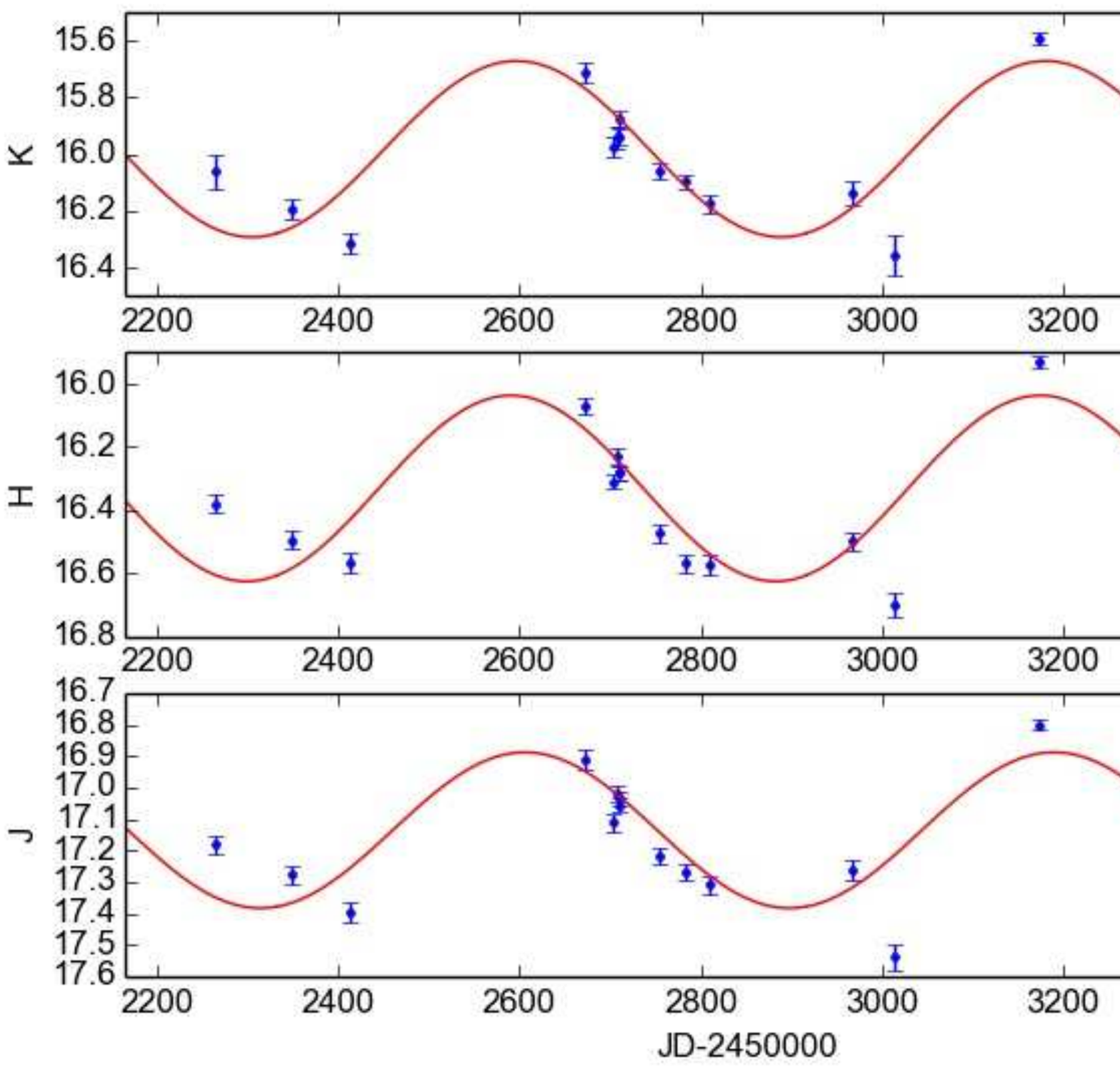} & \includegraphics[width=0.42\textwidth]{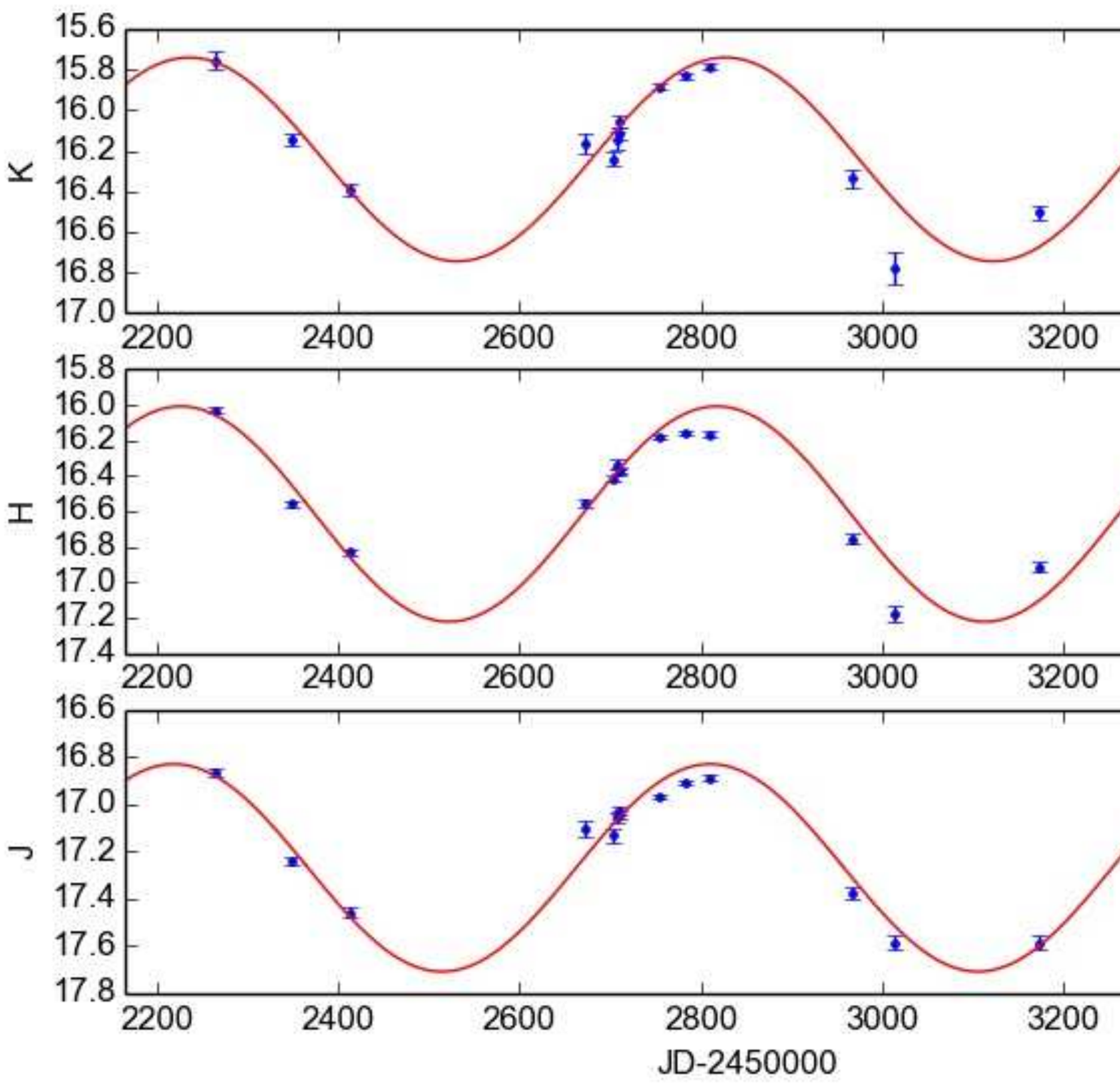}\\
(c) 1112 (583 d) & (d) 1067 (591 d)   \\
[-2pt]
 \includegraphics[width=0.42\textwidth]{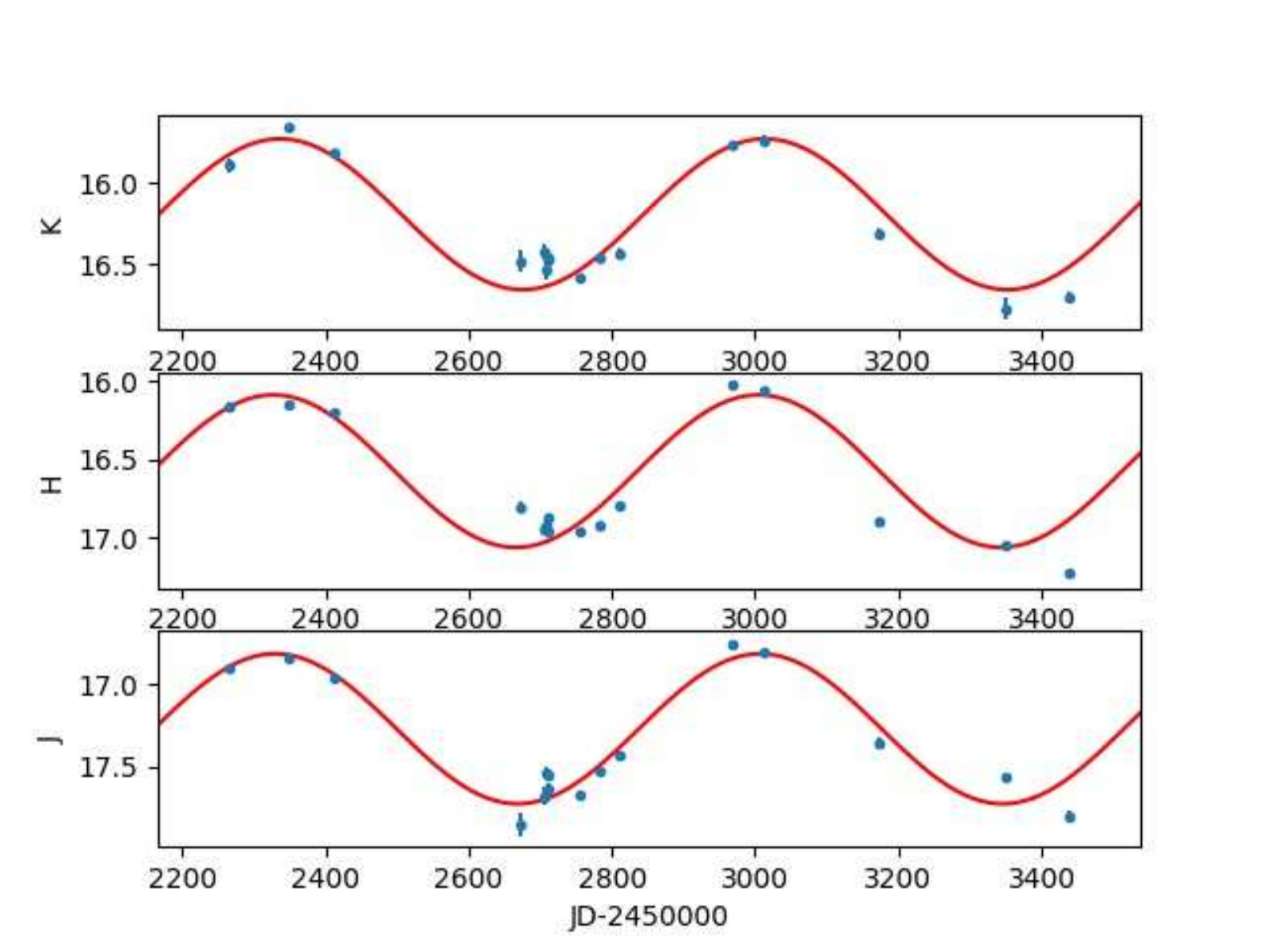} & \includegraphics[width=0.42\textwidth]{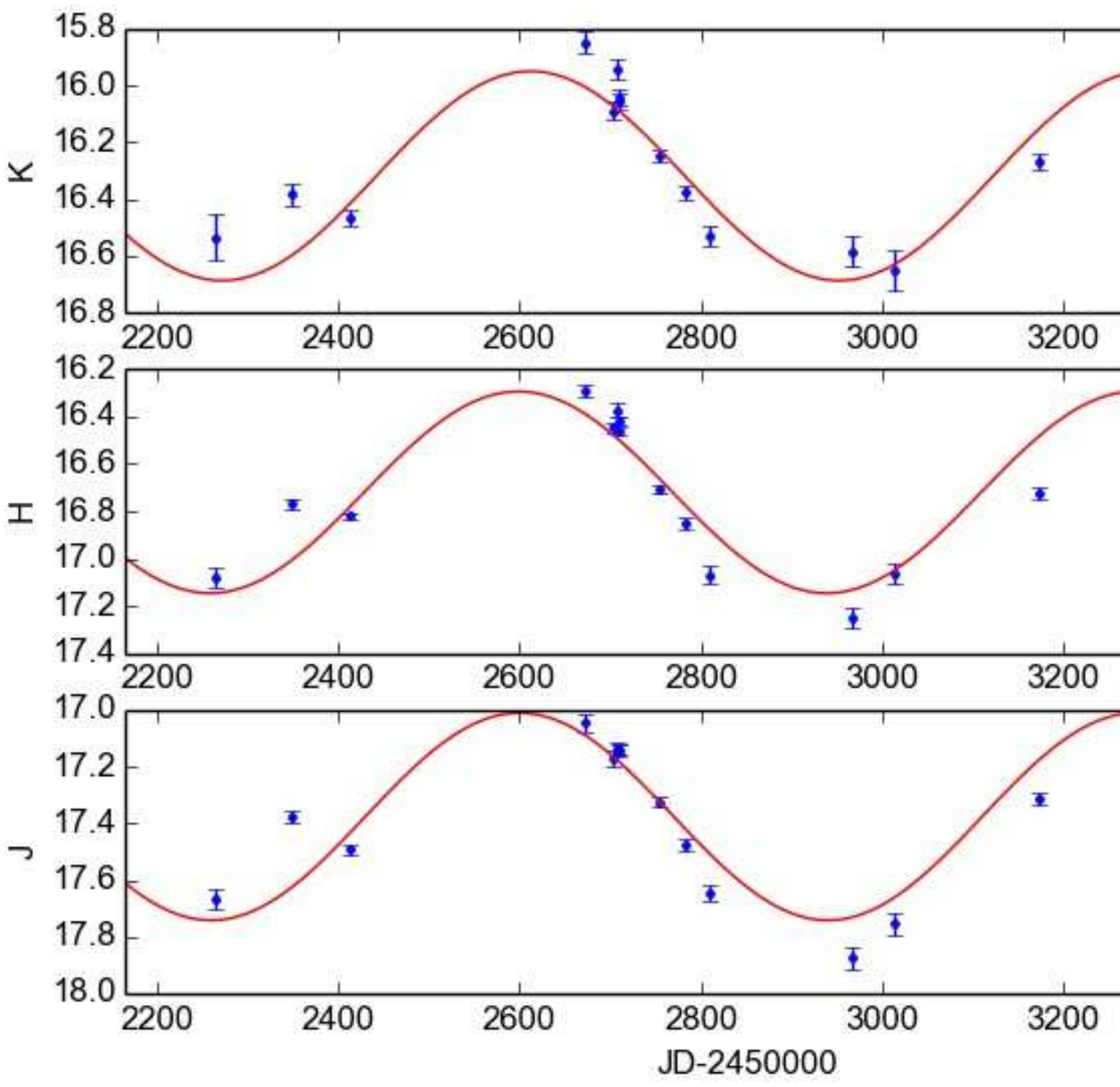} \\
(e) 1153 (678 d) & (f) 1224 (680 d)   \\
[-2pt]
\includegraphics[width=0.42\textwidth]{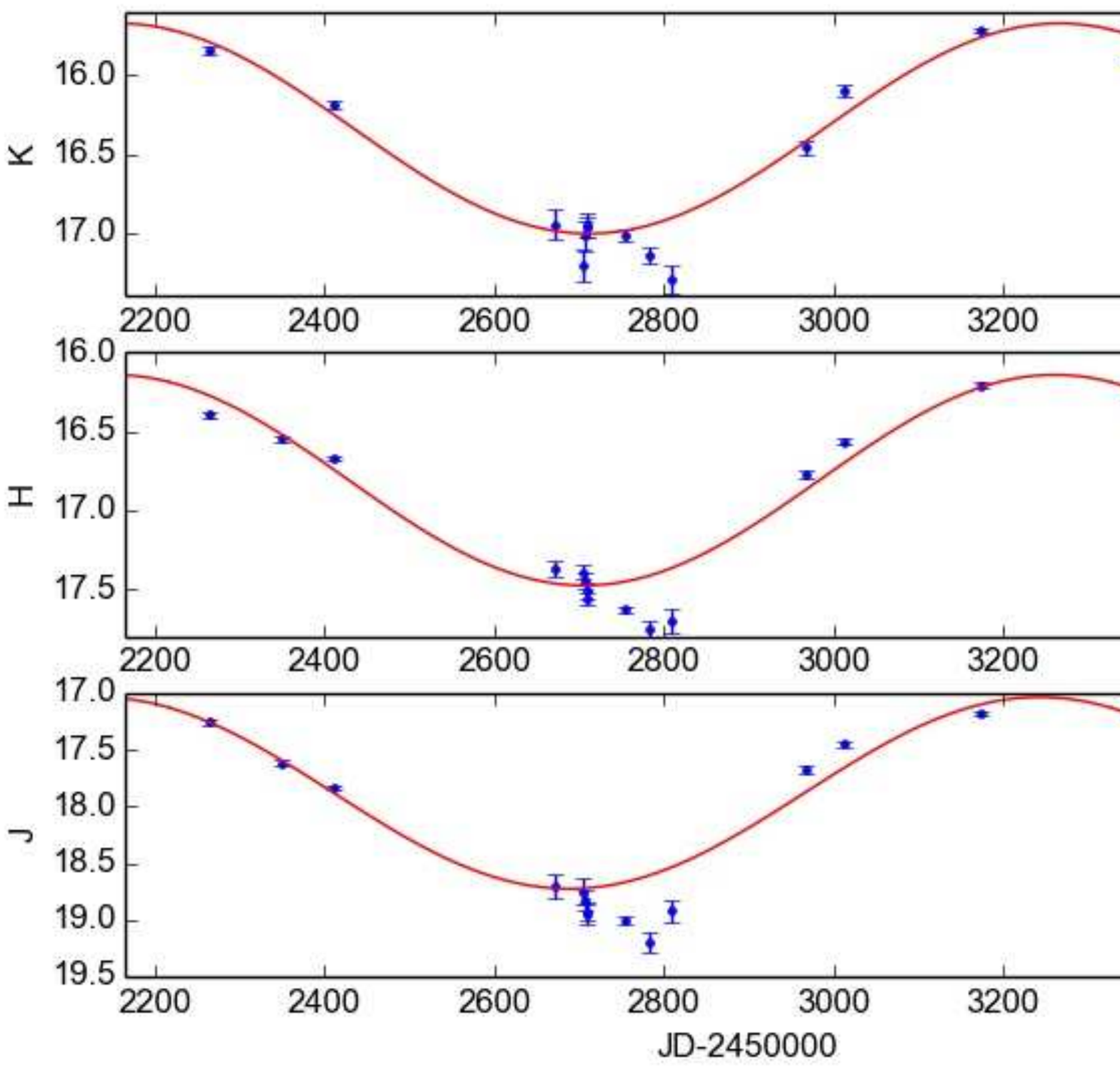} &\includegraphics[width=0.42\textwidth]{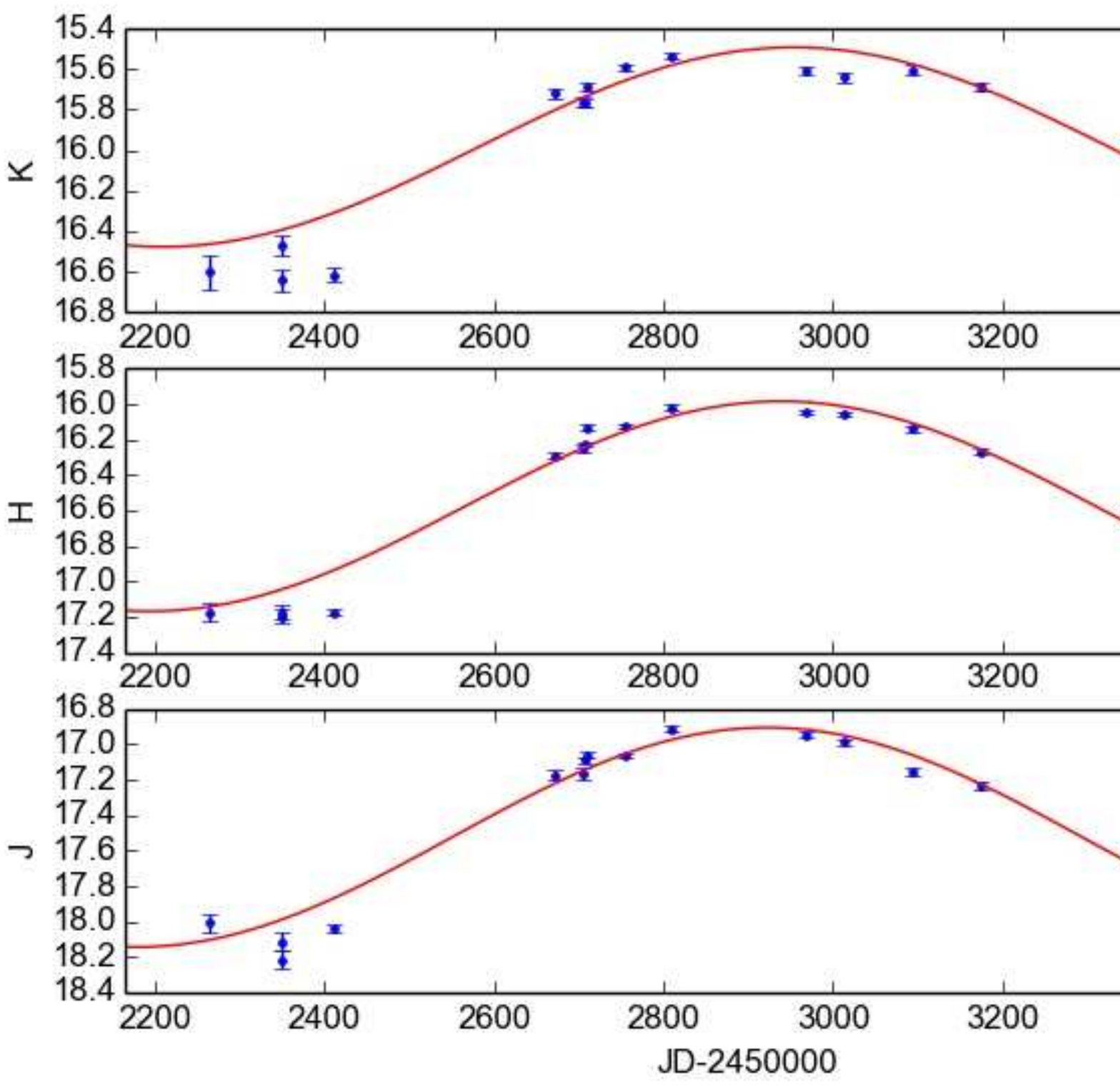} \\
(g) 2075 (1109 d) & (h)  3089 (1486 d) \\
[-2pt]
\end{tabular}
\caption{$JHK_S$ light curves for eight long period variables. Periods are shown in brackets following the star number. Curves are single-period sinusoidal fits to the data.}
\label{fig_SAAOvar}
\end{figure*}

The light curves  of the eight periodic variables of the eight periodic variables  are shown in Fig.~\ref{fig_SAAOvar} with single-period sinusoids overplotted on the observations. Mean magnitudes estimated from the lightcurve fits are listed, together with the derived periods and peak-to-valley amplitudes ($\Delta J, \Delta H, \Delta K_S$), in Table~\ref{tab_N3109var}.

\subsection{Variable object \,  \#1060}\label{ssec:3}
\begin{figure}
\includegraphics[width=0.47\textwidth]{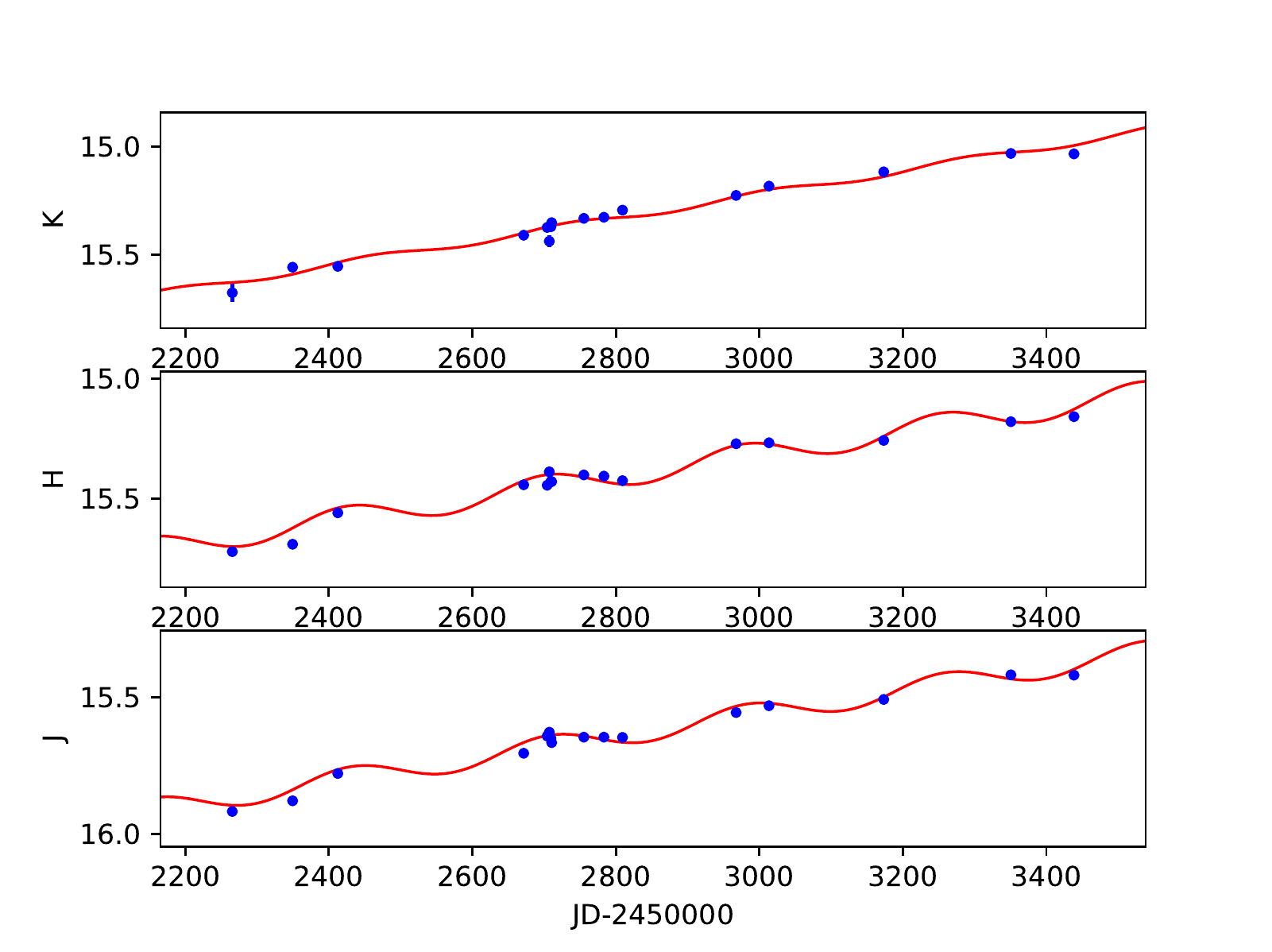}
	\caption{$JHK_S$ Lightcurves superimposed on the data for \#1060.}
	\label{LC_1060}
\end{figure}

In the case of object \, \#1060, the light curves show a steadily brightening trend with time. The $J$, $H$, $K_S$ magnitudes changed by $-0.64, -0.55$ and $-0.49$, respectively, over the period covered by our data. After linear slopes were removed, the period-finding program found a period of 276 day would fit the data, though the $K_S$ data are consistent with no variation. The data for this object are shown in Fig.~\ref{LC_1060}, together with the fitted two-component lightcurves. 

This variable is clearly blue on the basis of our photometry ($J-K_S=0.33, K_S=15.30$).
It is very close to the position of the 2MASS object, 2MASS\_J10030205--2608575, which is distinctly redder and fainter (2MASS: $J-K_S=0.77, K_S=15.78$), though the 2MASS errors are relatively large. The 2MASS star is identified as a red supergiant by \citet{Davidge_2018}.

 An object at the position of \#1060 was found in the Catalina survey to be variable, where it is designated \mbox{SSS\_J100302.1--260857},
 and is classified there as type VAR with a V magnitude range of 14.5 to 16.5.
 It appears in the Pan-STARRS1 survey DR1 \citep{Chambers_2016} as PSO J150.75885--26.1494 where it has clearly blue colours at epoch JD2456096.5, but is  
 flagged as extended.

 Our measurement appears to refer to the combined light of the red supergiant and another object. The 276-day variation might be due to the supergiant, but assuming that to have not varied since the epoch  
 of the 2MASS measurement, the contribution of the second object at our mid observational epoch would have been ($J-H=-0.15$, $H-K_S=-0.03$, $J-K_S=-0.19$, $K_S=16.43$) but the 2MASS errors are too large to attach any real significance to this result, save to say that the object is bluer than the supergiant. Taken at face value these colours suggest a significant emission line flux from the blue object.

 In the ACS Nearby Galaxy Survey Treasury images of NGC~3109 \citep{Holtzman_2006}, there is an object about 1.15 arcsec north-east of the supergiant; it is object \#12829 in Field 3, which is clearly blue with $F606W-F814W = 0.1$, corresponding to $V-I = 0.159$. Our mean epoch of observation was JD2452801, whereas
  that of the HST observation was JD2454411, so the object continued to be blue for some years after our observations. It could possibly be a background quasar or AGN that has undergone a long-lived outburst.

\subsection{C stars}\label{sec:Cstars}
Intrinsic carbon stars form following third dredge-up (3DUP) when
sufficient carbon reaches the stellar surface to change the surface
chemistry from C/O $<$ 1 to C/O $>$ 1 and therefore to change the dominant
molecules from $\rm H_2O, TiO$ etc to CN, CH, $\rm C_2$ etc. (note that
CO is always present and uses up most of either C (in O-rich stars) or O
(in C-rich) stars). The amount of 3DUP depends on the mass and
metallicity of the star, and the amount of 3DUP required to change the
surface chemistry depends on the initial oxygen abundance, as the more
oxygen initially present the more carbon is required to move the ratio
to C/O $>$ 1.  The relative numbers of C- and O-rich AGB stars is
theoretically a sensitive probe of mass and metallicity \citep[see,
e.g.][fig. 10]{Marigo_2017}, although it has yet to be calibrated against observations over
large ranges of these quantities.

The relative numbers of C stars and M stars is often used as a proxy for
metallicity and methods such as that devised by \citet{Cioni_2006} are
used to establish the region in a colour-magnitude diagram
\citep[e.g.][]{Jones_2018} where the O- and C-rich stars should be
expected. This is illustrated in Fig.~\ref{fig_SAAOphot_Tri} where the area in which O-rich
stars are anticipated at $\rm[Fe/H] = -1.2$  is shown as parallel dashed lines
\citep{Cioni_2006}. This is a representative metallicity for the AGB stars in NGC 3109 (see Fig.~\ref{fig:isochrones}); a change of $\pm$0.5 in [Fe/H] moves the lines by $\pm0.06$ in $J-K_S$. O-rich stars should fall between these lines, while
C stars should fall to the right.  Note that if this method was used on
NGC~3109 then almost one third of the C-stars we observed would have been
classified as O-rich. Thus this is not an effective way of isolating
C stars in galaxies that are significantly different from the Magellanic
Clouds, where the method was calibrated. In contrast, the Marigo models discussed in section~\ref{sec:LPMiras}  fit our observations of C stars rather well.

None of our variables  matches any star in the C star list of \citet{Demers_2003}. None of the 45 C stars in this list that we have measured is a large amplitude variable.
Our limiting magnitude is rather bright, being set by the faintest measurable stars on the reference $H$ image, so we have missed most of the C stars in our field. Close inspection of the $K_S$ images showed that a further 172 C stars were visible on at least 10 frames. We used Sextractor \citep{Bertin_1996} to measure these stars on as many frames as possible. The precision of the photometry is low ($\pm~0.2$ mag per point), but we found no evidence for large amplitude variability. On the other hand, as suggested later in section \ref{sec_evol}, variable star \#2075 may be an unrecognised C star (see section~\ref{sec_evol}). 

\subsection{Comparison with NGC~6822, WLM, IC~1613 and the SMC}

\begin{figure*}
\begin{tabular}{cc}
 \includegraphics[width=0.50\textwidth]{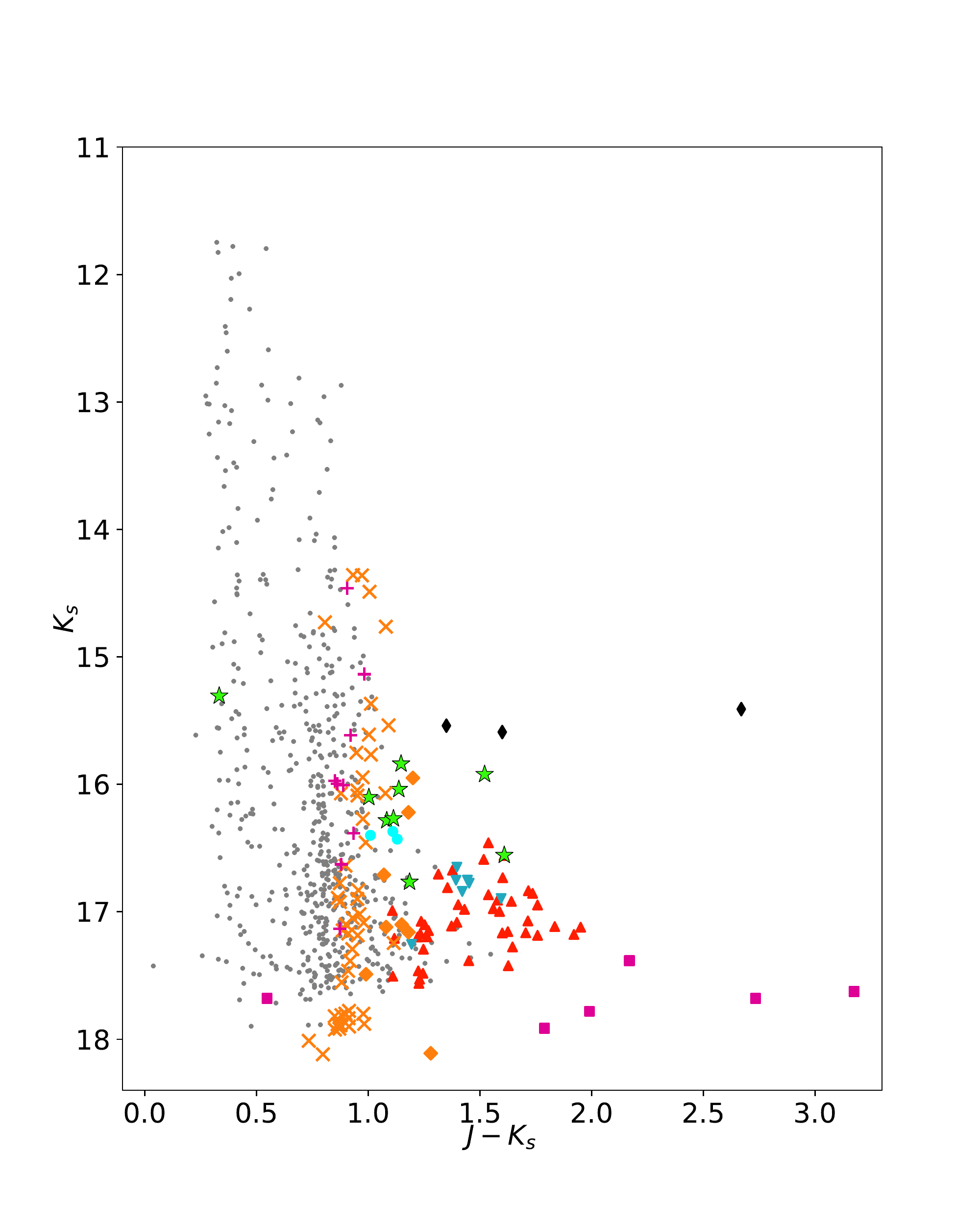} &   \includegraphics[width=0.50\textwidth]{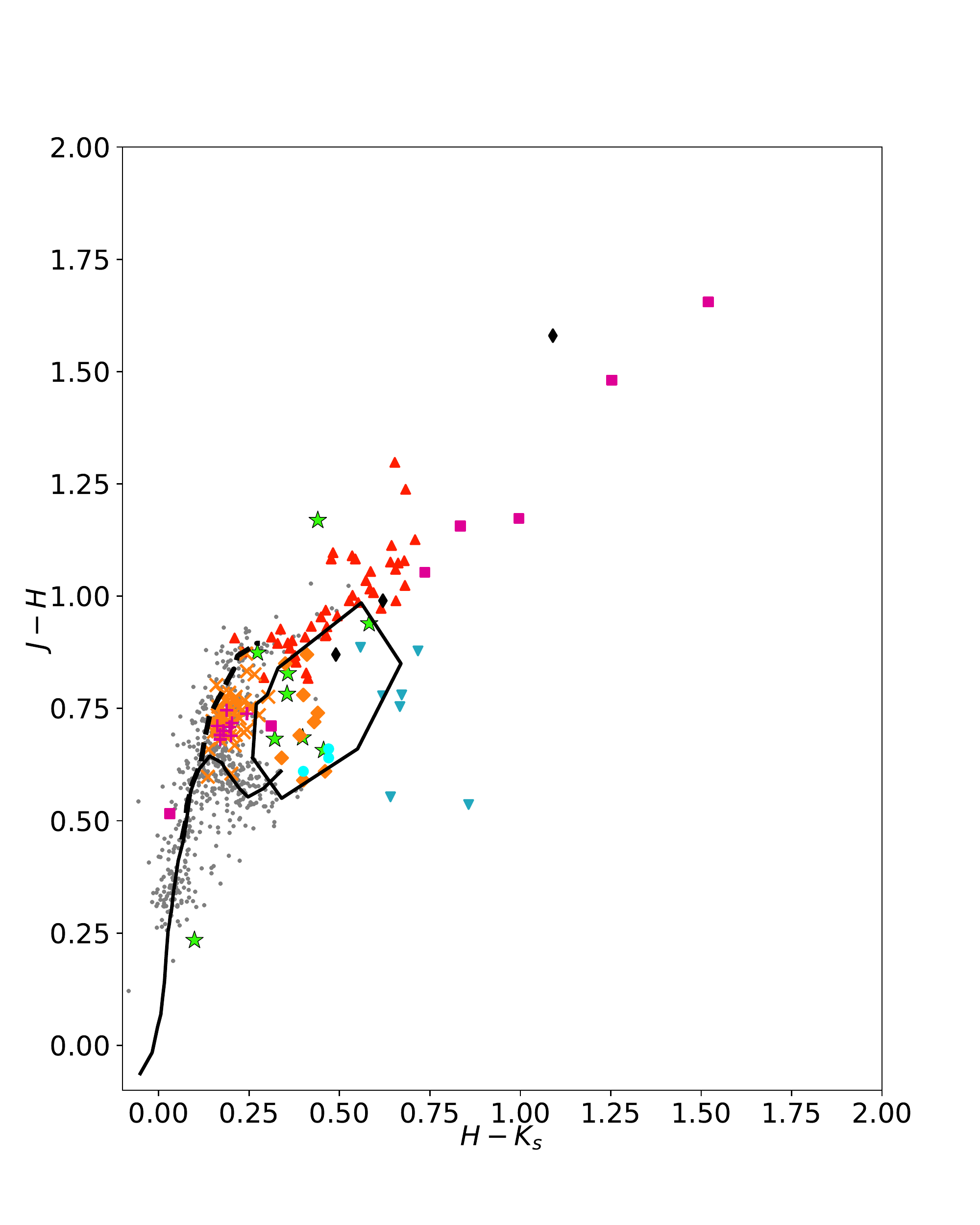} \\
 \end{tabular}
		\caption{Colour-magnitude (left) and two-colour (right) diagrams for NGC~3109 with points for NGC~6822, WLM, IC~1613 and the SMC added. All magnitudes have been shifted to the distance modulus of NGC~3109. 
	The X symbols (orange) show the positions of the M-type supergiants from \citet{Levesque_2012} and the diamond symbols (orange) the O-Miras from \citet{Whitelock_2013}, in the galaxy NGC~6822 adjusted for  reddening. The plus symbols and squares (magenta) show the  supergiants and variables, respectively, for WLM.  The filled circles (cyan) show the IC~1613 variables. The three SMC variables are shown as thin diamonds (black). Symbols and colours otherwise as in Fig.~\ref{fig_SAAOphot}.}
\label{fig_SAAOphot_comp}
\end{figure*}

It is instructive to compare the variables in NGC~3109 with those in other dwarf irregular galaxies, namely, NGC~6822,  WLM, IC~1613 and the SMC. In Fig.~\ref{fig_SAAOphot_comp} we have added our photometry for the red supergiants from \citet{Levesque_2012} and for the O-Miras from \citet{Whitelock_2013} in NGC~6822, our unpublished data for the red supergiants \citep{Levesque_2012} and newly discovered variables in WLM (Menzies, in preparation), as well as for the four O-rich Miras in IC~1613 \citep{Menzies_2015}. We have corrected the NGC 6822 data for the differential distance modulus of 2.03 mag and an assumed reddening for that galaxy of E($J-K_S$) = 0.13 mag. For WLM and IC~1613, where the reddening is negligible, the $K_S$ magnitudes have been adjusted downward by 0.7 and 1.42 mag, respectively. Data for the SMC variables are taken from the discussion in section~\ref{sec:LPMiras}.
The five variables in NGC~3109 with periods between 430 and 680 days lie to the red of the supergiants in the colour-magnitude and two-colour diagrams. They occupy the same regions as do the HBB Miras in NGC~68222 and IC~1613. 

\subsection{Period-Luminosity Relations}
\begin{figure}
\includegraphics[width=8.5cm]{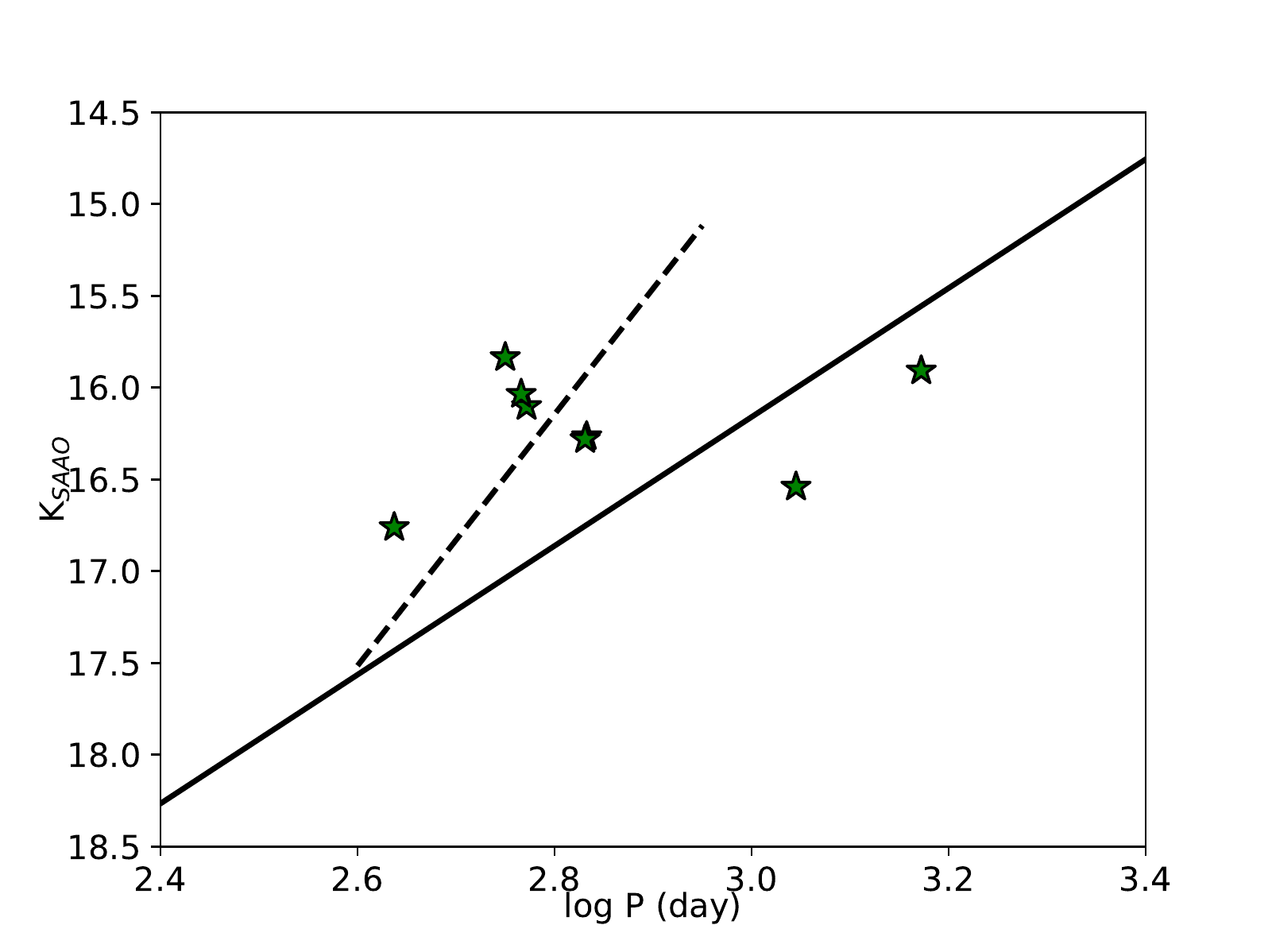}
	\caption{Period-luminosity diagram for NGC~3109 Mira variables (green $\ast$ symbols). 
	The solid line shows the relation for the LMC \citet{Whitelock2008} with an assumed distance modulus of 25.6 for NGC~3109%
 while the dashed line %
shows the PL relation of \citet{Ita_2011} for stars with log P $>$ 2.6. Note that the K magnitudes are on the SAAO system, not the 2MASS one.}
\label{PL_diag}
\end{figure}

The log period-apparent $K_S$ luminosity diagram based on the data from Table~\ref{tab_N3109var} 
is compared with the PL relation for the LMC \citep{Whitelock2008} shifted from a distance modulus of 18.5 to our assumed value of 25.6 for NGC~3109 is shown in Fig.~\ref{PL_diag}. The linear relation was established by \citet{Whitelock2008} for log P $\le$ 2.6 and is extrapolated beyond that in the diagram. The C-Miras follow the same relation up to log P = 2.6. We include the period-luminosity relation found by \citet{Ita_2011} for variables with log P $>$ 2.6. 

While the two reddest variables sit close to the LMC line, the other six are significantly brighter. They are probably O-rich stars and at least five of them are undergoing hot-bottom burning. 

\begin{figure}
\includegraphics[width=8.5cm]{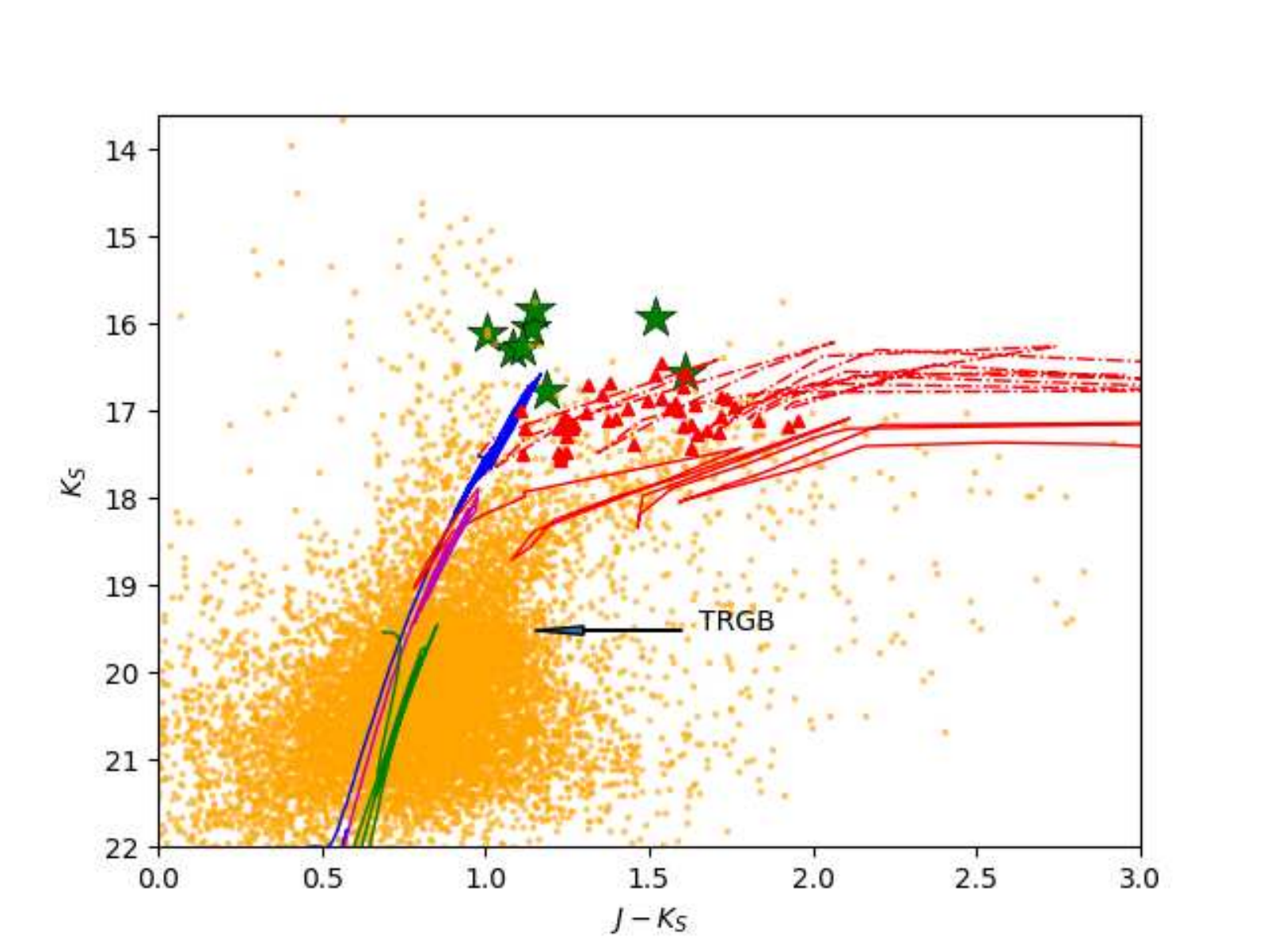}
	\caption{Comparison of isochrones with the observed colour-magnitude diagram. The background points are from \citet{Gorski_2011}, the $\ast$ symbols (green) are our variables, and the (red) triangles show the C stars we measured in the present study.  The near-vertical tracks are isochrones for, from left to right, ages of 0.398 (blue), 0.798 (magenta) and 10.0 Gyr (green), respectively. The isochrones have different metallicities, decreasing from left to right, of Z = 0.001, 0.0007, 0.0003, respectively. The dash-dot (red) line is the continuation of the 0.398 Gyr isochrone after stars become C-rich, while the continuous (red) line is the same for the 0.798 Gyr isochrone. Note that the tip of the 10 Gyr isochrone coincides with the tip of the red giant branch as measured by Gorsk\'i.} 
\label{fig:isochrones}
\end{figure}

\section{Evolutionary Status of the Variables and C Stars}\label{sec_evol}
Star formation in NGC~3109 has taken place in two episodes \citep{Weisz_2011}. In the first, almost 80\% of the galaxy's stars were formed by about 10 Gyr ago; these presumably are the giants and fainter stars with a very low metallicity. In more recent times, since about 2 Gyr ago, stars have been forming at a slow rate with presumably a steady increase in metallicity to the level seen in the blue supergiants \citep[e.g.][]{Tramper_2014}. Thus we would expect there to be a range of metallicity and age amongst the AGB stars. To illustrate this, we have obtained a selection of the latest \textsc{parsec + colibri} tracks \citep{Marigo_2017} and overlaid them on the colour-magnitude diagram of \citet{Gorski_2011} in Fig.~\ref{fig:isochrones}. This can only be illustrative, but suggests a plausible range of ages and metallicities for the AGB of NGC~3109. Note that the 0.398 Gyr track covers the C stars that we measured, while the 0.798 Gyr track overlies the bulk of the fainter, red AGB stars covered by Gorsk\'i's measurements.

\begin{figure}
\includegraphics[width=8.5cm]{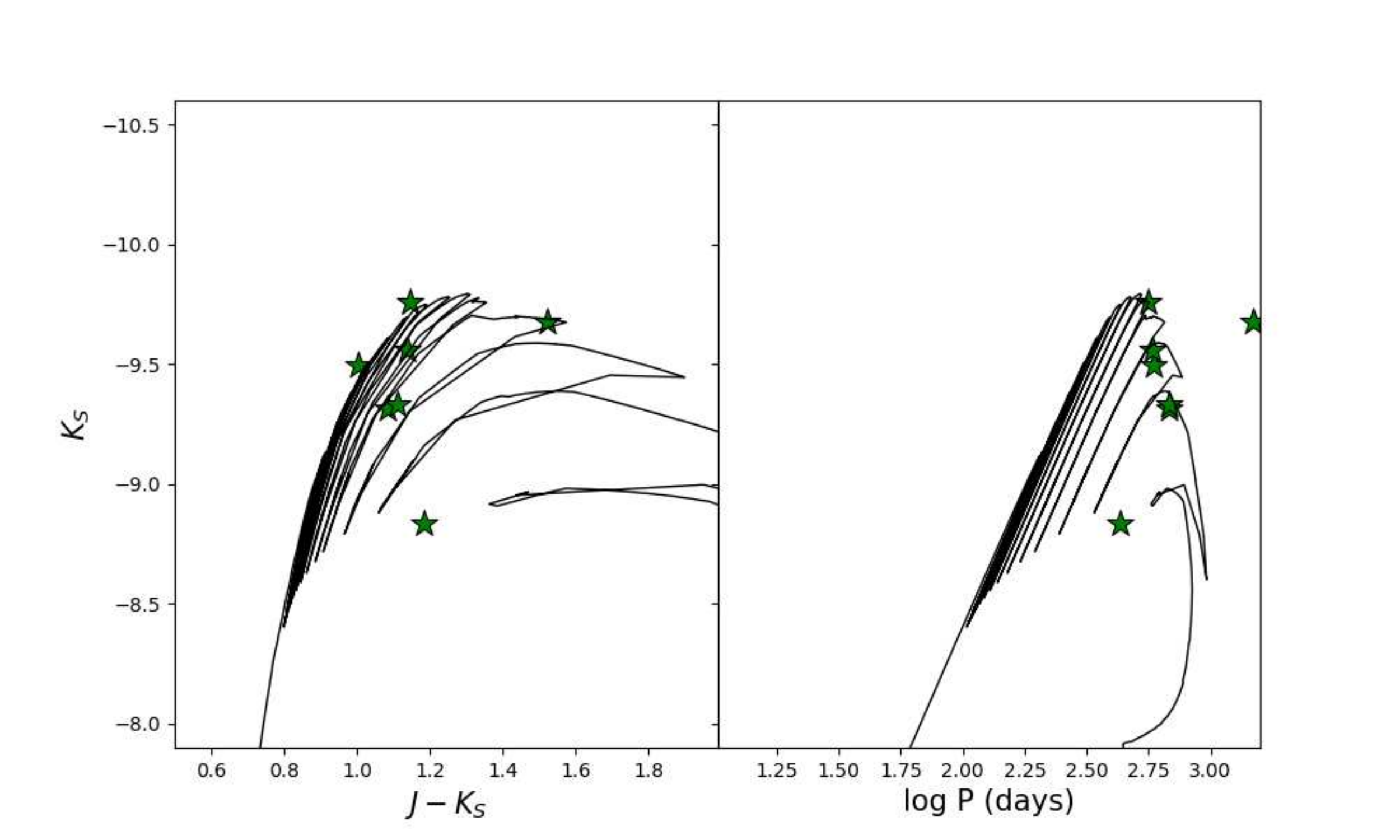}
	\caption{Evolutionary track (left panel) and period evolution (right panel) for a 5 M$_\odot$star with Z=0.0005. The variables from this paper except for \#2075 are shown as $\ast$.}
\label{fig:HBB}
\end{figure}

To investigate the nature of the variables further we have used a small set of evolutionary tracks for the thermally-pulsing phase of the AGB \citep{Marigo_2017} (kindly supplied by Paola Marigo)
covering the mass range from 2.6 M$_\odot$ to 5 M$_\odot$, for metallicities Z=(0.0005, 0.001, 0.004). Five of the six proposed HBB variables are well represented by a track for a star with an O-rich atmosphere, M=5 M$_\odot$ and Z=0.0005 to 0.001. This is in the mass range expected for HBB stars \citep[see, eg.][]{Marigo_2017}. The predicted periods at the luminosities of these variables are good matches to those observed. The track and the period evolution are shown in Fig.~\ref{fig:HBB}.
The sixth variable (\#2081), which is the faintest, shortest period one, appears to lie on a track (not illustrated here) for a 3--4 M$_\odot$, Z=0.001--0.004 O-rich star where again the predicted period matches the observed one. It could also be a $\sim$3 M$_\odot$ star on a track soon after becoming a C star. In the two-colour diagram, Fig~\ref{fig_SAAOphot}, it is outside the O-Mira box, and at the blue end of the C-star region, though it is not identified as a C star by \citet{Demers_2003}. It may be in a transitional evolutionary phase. 
The longest period star (\#3089)  can be fitted by O-rich tracks for a 5 M$_\odot$ star with Z in the range 0.0005--0.004. Its observed period far exceeds the maximum predicted for any of these tracks.

In the case of star \#2075 the situation is ambiguous. It appears to be best explained as a C-rich star with mass $\sim$3 M$_\odot$ and Z = 0.0005--0.004,  although the observed period is only predicted by the Z=0.004 track. This interpretation is supported by the star's position in the colour-colour diagram, Fig.~\ref{fig_SAAOphot}. On the other hand, it could be an O-rich star of mass 4--5 M$_\odot$, and metallicity Z=0.001--0.004, though only for the Z=0.004 case is the period predicted correctly.
The variable is not listed as a C star by \citet{Demers_2003} but has a large amplitude. It falls in a gap between CCDs in the  CFH12K camera on the Canada-France-Hawaii Telescope, so was missed in the C star survey (S. Demers, private communication). 

Further discussion of the nature of the variables can be found in section~\ref{sec_class} in the Appendix. \citet{Lebzelter_2018} have proposed a method of separating AGB stars into different groups on the basis of a combination of \textit{Gaia} and near-infrared photometry, which we have applied to variables in NGC~3109 and in the LMC, the SMC and some of the Local Group galaxies that we have observed in the past. For NGC~3109, we find general agreement between the classification based on this approach and that based on the evolutionary tracks.

\section {Very Long Period Miras}\label{sec:LPMiras}
There are only a few large amplitude AGB variables known with primary pulsation periods over 1000 days, partly because they are rare, but the limited patience of observers and TACs also makes their characterization difficult. These stars are of special interest because they are uncommon and will include short-lived evolutionary phases of the most massive stars that experience the AGB. In our paper on Sgr~dIG \citep{Whitelock_2018} we discussed a Mira with a period of $P=950$ days and suggested it was in a post-HBB phase and leaving the AGB. In view of the fact that we have found two even longer period stars in NGC3109, it is appropriate to briefly review how many very long period stars there are in the Local Group and what we know about their evolution. We start with a brief census of $P>1000$ day Miras in the LMC, SMC and the Galaxy.

{\bf LMC:} \citet{Wood_1992} measured periods of over 1000 days for eight IRAS sources in the LMC and detected OH emission in four of them, confirming their O-rich nature. \citet{Whitelock_2003} confirmed the long periods for six of these and measured another three with $P>1000$ days. OGLE \citep{Soszynski_2009} confirmed the long periods for seven of the large amplitude variables and identified one more\footnote{The OGLE catalogue gives multiple periods for many stars. Here we only consider the dominant one.}. On the basis of their colours \citet{Soszynski_2009} suggested that seven of the variables with long periods were O-rich and one, IRAS05402--6956 (P=1352 day), was C-rich. However, this cannot be a C-star as it was one of those that \citet{Wood_1992} detected as an OH maser, a detection that has recently been confirmed \citep{Goldman_2017}. \citet{Groenewegen_2018} added two further O-rich sources and five C-rich sources, as well as refining some of the periods. So the total census for the LMC is 13 O-rich large amplitude variables with periods up to 1362 days and five C-rich sources with periods up to 1209 days; these are listed in Table~\ref{tab_LMC}

\begin{table}
\caption{Very Long Period Variables in the LMC}
\centering
\begin{tabular}{lccc}
Name & RA & Dec & Period \\
	& \multicolumn{2}{c}{(2000.0)}&  (day) \\
\multicolumn{4}{l}{Carbon rich}\\
 IRAS 05125--7035  &078.003208 & -70.540000  & 1115 \\
  IRAS 05278--6942  &081.850458  &-69.662472   &1001  \\
 IRAS 05506--7053    &087.485500  &-70.886583  & 1026  \\
  IRAS 05509--6956   &087.608770  &-69.934250 &  1052  \\
 IRAS 05568--6753   &089.161500  &-67.892889  & 1209 \\
\multicolumn{4}{l}{Oxygen rich}\\
IRAS 04407--7000 &  070.118667 & -69.920417 &  1148  \\
IRAS 04498--6842  &072.422792  &-68.630944  & 1256  \\
IRAS 04509--6922 & 072.668583  &-69.292194  & 1271 \\
IRAS 04516--6902 &072.870792  &-68.963889  & 1095  \\
 IRAS 04545--7000 & 073.541875 & -69.932833 &  1254 \\
MSX LMC 1210      &073.835417  &-68.377583  & 1050  \\
HV 888                 &076.058875  &-67.270639  & 1005  \\
MSX LMC 642       &082.200708 & -71.041361  & 1133  \\
IRAS 05298--6957  & 082.352600 & -69.920468  & 1265   \\
MSX LMC 807         & 083.154833  &-67.115667  & 1069  \\
IRAS 05329--6708  &083.213875  &-67.114444  & 1303 \\
IRAS 05402--6956   &084.937000  &-69.921667  & 1362 \\
IRAS 05558--7000  & 088.837625 & -70.000833  & 1175    \\ 
\end{tabular}
\label{tab_LMC}
\end{table}

\begin{table*}
\caption{SAAO $JHKL$ Photometry of SMC Variables (all O-rich)}
\begin{threeparttable}
\centering
\begin{tabular}{ccccccccccccc}

	Name & RA & Dec & Period & $\Delta J$ & $\Delta H$ & $\Delta K$ & $K$ & $J-H$ & $H-K$ &$ J-K$ &$K_S$\tnote{*} & $J-K_S$\tnote{*}  \\
	& \multicolumn{2}{c}{(2000.0)}&  day   & \multicolumn{9}{c}{ mag}\\
HV11417 & 015.20071  & -72.85056  &  1092  & 1.26 & 1.48 & 1.50 & 8.86 &0.97 & 0.47 & 1.44 & 8.86 & 1.35 \\
GM103 & 012.62764 & -72.85830 &1062 & 1.27 & 1.34 & 1.18& 8.91 & 1.10 & 0.61 & 1.70 & 8.91 & 1.60\\
IRAS00483--7347&012.52963 &-73.52377&1859&0.98 &0.79 & 0.57 & 8.71& 1.72 & 1.11 & 2.83 & 8.73 & 2.67\\
\end{tabular}
\begin{tablenotes}
  \small
  \item[*] Transformed from the SAAO to the 2MASS system.
\end{tablenotes}
\end{threeparttable}       
\label{tab_SMC}
\end{table*}
	
{\bf SMC:} The OGLE group \citep{Soszynski_2011} measured periods for three O-rich Miras in the SMC of  between 1062  and 1859 days.  All three Miras have confirmed M-type spectra \citep{Elias1980, Groenewegen_1998, vanLoon2008}.
	$JHKL$ photometry for these three stars is given in Appendix~\ref{sec_LPV_SMC}  and a summary of their mean colours and amplitudes is given in Table~\ref{tab_SMC}, where the periods are taken from OGLE \citep{Soszynski_2011}, but agree with what is derived from the measurements in the Appendix.  
	
	HV\,11417 is sometimes classed as a supergiant, but these colours put it in the region of the two-colour diagram occupied by Miras \citep{Feast_1984} which, together with its large amplitude, supports the AGB classification favoured recently \citep{Soszynski_2011, Groenewegen_2018}. The other two stars have the colours of Miras reddened by circumstellar reddening and could be either O- or C-rich. All three stars are shown in Fig.~\ref{gaia_diag_sg} where IRAS00483--7347 falls in the region dominated by C stars, showing that the classification scheme based on \textit{Gaia} and near infrared photometry breaks down (as do other photometrically-based schemes) for O-rich stars with high mass-loss rates. Transforming the photometry to the 2MASS system following \citet{Carpenter_2001}\footnote{ updated: http://www.astro.caltech.edu/$\sim$jmc/2mass/v3/transformations/} gives the values listed in the last two columns of the table. These are shown in Fig.~\ref{gaia_diag_sg} for comparison with the NGC~3109 Miras, assuming a distance modulus of 18.9 for the SMC. 

\citet{Groenewegen_2018} suggest that IRAS 00483--7347 (MSX SMC 055)  is
the best Magellanic Cloud candidate for a super-AGB star, based on its
high luminosity ($M_{bol}=-8.0$), long period and large amplitude. They
propose a current mass of $M=8.5\pm1.6 M_{\odot}$ and an initial mass
around $1M_{\odot}$ larger than this. A high mass is independently
supported by the measured high rubidium abundance \citep{Garcia_2009}.

{\bf The Galaxy:} The GCVS \citep{Samus_2017} lists 11 Miras with periods over 1000 (up to 1994) days. Only one of these, V829~Cas, with a period of 1060 days \citep{Groenewegen_1998}, is a C star \citep{Zuckerman_Dyck_1986}.  This is the longest period known for a C-rich Mira in the Galaxy, and the only one above 1000 days. The models discussed by \citet{Marigo_2017}  do not produce C stars at high metallicities,  so we would anticipate that such stars are less common in the Galaxy, particularly among the higher-mass longer-period AGB stars, than e.g., in the Magellanic Clouds.

Four of the other ten GCVS variables are either not Miras or have periods much shorter than 1000 days. The ASAS online catalogue \citep{Pojmanski_1997} convincingly provides shorter periods for three of them,  CD Pup, V1156 Sgr and V581 CrA (the last is actually an RCB star \citep{Miller_2012}) and the AAVSO catalogue \citep{Watson_2006} gives a short period for EI~Sct.  The other six long period Miras in the GCVS are OH/IR stars, so they are definitely AGB stars and their long periods are at least plausible; these are listed in Table~\ref{lpv} with their periods and the uncertainty on the period ($\Delta P$) where it is given in the reference. The GCVS also lists three possible Miras with $P>1000$ days; one of these, AY Lac, is a cataclysmic variable \citep{Kroll_2003}, the second, V669~Cas, is a well known OH/IR star and is included in the Table. Very little is known about the third, CP Sco, but the AAVSO list a short period, so it is not included here. 

There are another nine OH/IR sources for which periods over 1000 days have been determined, listed in Table~\ref{lpv}, and several of them already have GCVS designations. So there are at least 16 O-rich Miras with periods over 1000 days in the Galaxy.

In addition to the variables discussed above there are several other OH-variables within the Galaxy with periods over 1000 days (a few over 2000 days) \citep[e.g.,][]{Herman_Habing_1985, van_Langevelde_1990},  most of which are probably massive AGB stars (some may be supergiants), but have no optical or near-infrared counterpart, presumably because of circumstellar and/or interstellar absorption (they are close to the Galactic plane).  So, in summary, within the Galaxy there is one C star and at least 16 O-rich stars with periods over 1000 days. The census is obviously incomplete, but the total number cannot be very large. 

Long periods among Miras are rare, either because they are in a very short-lived evolutionary phase \citep[see, e.g.][]{Marigo_2017} or because they only occur among the most massive, and therefore the rarest, of  the AGB stars, or both.  As summarized by \citet{Feast_2009}, we understand that the initial mass of Miras with periods less than about 400 days is a function of their pulsation period. The situation is less clear for longer period stars, some of which experience HBB \citep[see, e.g.][]{Whitelock_2017, Whitelock_2018}. Very long period OH-variables, which can be detected to very large distances, are found only close to the Galactic plane, supporting the view that they have progenitors that are significantly more massive than shorter period Miras. 

It is therefore very interesting that we find two very long period stars in NGC~3109.  If \#2075 is indeed a C star (section 5), its period is comparable to those of the longest period LMC C-rich Miras and longer than any C-Mira known in the Galaxy. In view of the problems there have been in classifying LMC  long-period stars as O- or C-rich, as described above, this must be confirmed using spectroscopy or narrow band filters. The period of 1486 days for \#3089 is somewhat uncertain because of the limited temporal coverage, but the period cannot be under 1000 days. Both of these stars are much bluer (lower $J-K_S$) than Miras of comparable period in the Galaxy or the LMC, indicating lower circumstellar extinction and weaker stellar winds.  \citet{Whitelock_2018} discussed a long period ($\sim 950$ day) O-rich Mira in Sgr~dIG, emphasizing how unusual it was to find such a long period Mira with such blue colours. Using stellar evolution models they suggested that its progenitor had a mass around $5M_\odot$, and that it was close to the end of its AGB evolution. It is interesting that we find a very similar star in NGC~3109, which galaxy has about 20 times the mass in the form of stars than does Sgr~dIG \citep{McConnachie_2012} (there is insufficient information to comment on the relative number of stars of the appropriate age to produce long period Miras). It is not entirely clear what metallicity we should associate with these, presumably massive, AGB stars, but probably not as low as in Sgr dIG. The metallicity is important as it affects the mass-loss rates of O-rich stars much more strongly than those of C-rich ones \citep{Wood_1992, Matsuura_2005}. As mentioned above, there is strong evidence that the young stars in NGC~3109 have higher metallicities than do those on the giant branch.
	
	This is important in the context of the conclusions of \citet{Goldman_2018} who suggest that the dusty winds, which are generally found towards the end of the evolution of the most massive cool stars, are curtailed at low metallicity \citep[see also][and references therein]{Wood_1992, Matsuura_2005}.  It is also potentially important in the context of electron-capture supernova (ECSNe) \citep[][section 7.1 and references therein]{Langer_2012}, the progenitors of which are probably massive AGB stars. The frequency of ECSNe may depend on metallicity, if at low metallicity the stellar winds are weaker, so that evolution is not terminated by mass loss and the core has time to grow large enough for the star to become a supernova. The discovery of two luminous long period AGB variables in the metal weak environment of NGC~3109 lends support to this channel for ECSNe.

\begin{table}
\centering
\caption{Long period Miras in the Galaxy}
\begin{tabular}{lcrrl}
	\hline
	name & P & $\Delta P$ & Ref & other\\
	\hline
	V0829 Cas* & 1060& & 1 & IRAS01144+6658\\
	V1360 Aql & 1140 & & 1 & OH30.7+0.4\\
	V1362 Aql & 1730 & & 1 & OH30.1--0.7\\
	V1363 Aql & 1490 & & 1 & OH32.0--0.5 \\
	V1365 Aql & 1750 & & 1 &OH32.8--0.3\\
	V1366 Aql & 1340 & & 1 &OH39.7 +1.5\\
	V1368 Aql & 1650 & & 1 & OH 42.3--0.1\\
	V0669 Cas & 1537 &18 & 1,2 & OH127.8--0.0\\
	NSV25875 & 1460 & 11 & 3 & OH104.9+2.4\\
	IRAS03293+6010 & 1800 & 400 & 4 & OH141.7+3.5\\
	IRAS05131+4530 & 1100 & 100 & 4 & AFGL712\\
	IRAS07222--2005 & 1200 & 200 & 4,7 & \\
	V1185 Sco & 1400: & & 4 & OH357--1.3\\
	V437 Sct & 1589 & 42 & 5 & OH26.5+0.6 \\
	V438 Sct & 1172 & 94 & 5 & OH26.2--0.6\\
	V441 Sct & 1785 & 114 & 5 & OH21.5+0.5\\
	IRAS03206+6521 & 1410 & & 6 & OH138.0+7.2\\
	\hline

\end{tabular}

Notes: * C star, see text. 1 GCVS \citep{Samus_2017}; 2 \citet{Suh_Kim_2002}; 3 \citet{Riechers_2005}; 4 \citet{Jimenez_2006}; 5 \citet{van_Langevelde_1990}; 6 \citet{Orosz_2017}; 7 \citet{Perez_2017} detect OH and Rb.\\
\label{lpv}
\end{table}

\section*{Acknowledgements}

JWM and PAW wish to thank Prof Rolf Kudritski for providing access to the facilities of MIAPP, where part of the work on this paper was done.
This research has made use of Aladin \citep{Bonnarel_2000}.
This material is based upon work supported financially by the South African National Research Foundation.
Some of the data presented in this paper were obtained from the Mikulski Archive for Space Telescopes (MAST). STScI is operated by the Association of Universities for Research in Astronomy, Inc., under NASA contract NAS5-26555.
This publication makes use of data products from the Two Micron All Sky Survey, which is a joint project of the University of Massachusetts and the Infrared Processing and Analysis Center/California Institute of Technology, funded by the National Aeronautics and Space Administration and the National Science Foundation.
The IRSF project is a collaboration between Nagoya University and the SAAO supported by the Grants-in-Aid for Scientific Research on Priority Areas (A) (no. 10147207 and no. 10147214) and Optical \& Near-Infrared Astronomy Inter-University Cooperation Program, from the Ministry of Education, Culture, Sports, Science and Technology (MEXT) of Japan and the National Research Foundation (NRF) of South Africa.
This research has made use of the VizieR catalogue access tool, CDS, Strasbourg, France. The original description of the VizieR service was published in A\&AS 143, 23.
This work has made use of data from the European Space Agency (ESA) mission
\textit{Gaia} (\url{https://www.cosmos.esa.int/gaia}), processed by the \textit{Gaia}
Data Processing and Analysis Consortium (DPAC,
\url{https://www.cosmos.esa.int/web/gaia/dpac/consortium}). Funding for the DPAC
has been provided by national institutions, in particular the institutions
participating in the \textit{Gaia} Multilateral Agreement.
We are particularly grateful to Paola Marigo who generated evolutionary tracks for us. 
We thank Serge Demers for promptly providing a catalogue of C stars in the region of variable \#2075. 
We also thank Marek Gorsk\'i for providing the \textsc{araucaria} near-infrared data. 
For their contributions to the original IRSF observations, we wish to thank Toshihiko Tanab\'e, Yoshifusa Ita, Shogo Nishiyama, Yasuaki Haba and Enrico Olivier.
The following people contributed observations to the data in the Appendix: Enrico Olivier, Jacco van Loon, Albert Zijlstra.

\bibliographystyle{mnras}
\bibliography{N3109_bib.bib}
\appendix

\section{\textit{Gaia} Photometry and Classes of Variables}\label{sec_class}
\citet{Lebzelter_2018} have proposed an interesting method for separating AGB stars
	into different subgroups (e.g. C-rich, O-rich AGB stars of low mass, etc) based on \textit{Gaia} DR2\citep{Gaia_2016, Gaia_2018} $BP$ and $RP$ magnitudes combined with 2MASS $JK_S$ magnitudes. They considered the LMC and showed how stars grouped in the $K_S$, $W_{RP}-W_K$ plane, where $W_{RP}$ and  $W_K$ are Wesenheit functions \citep[see][for details]{Lebzelter_2018}.

	We constructed such a diagram for SMC AGB stars and found a very similar distribution to that for the LMC, in spite of the metallicity difference. Encouraged by this we computed the Wesenheit magnitudes for the long-period variables in NGC~3109, IC~1613, NGC~6822, the three very long period SMC variables from Table~\ref{tab_SMC} and the eight variables from Table~\ref{tab_sgGaia} that have \textit{Gaia} data. A matching radius of 1 arcsec was used in all cases.

\begin{table*}
	\caption{$JK_S$ and \textit{Gaia} ($BP$,$RP$) Photometry of LPV in LG Galaxies} 
\begin{threeparttable}
\centering
\begin{tabular}{cccccccc}
	Galaxy/Star \tnote{*}  & RA   & Dec   &  $K_S$  & $J-K_S$ & $RP$  & $BP-RP$   & $W_{RP}-W_K$        \\
	 & \multicolumn{2}{c}{(2000.0)}   &  &  &  &  &  \\
IC 1613  &    &    &    &    &    &    &      \\
1017  &  16.27831  &  2.22648  &  15.14  &  1.11  &  18.26  &  2.53  &  0.60   \\
3011  &  16.20428  &  2.09140  &  15.17  &  1.01  &  17.92  &  2.57  &  0.10   \\
2035  &  16.11180  &  2.17299  &  15.20  &  1.13  &  17.99  &  2.15  &  0.77   \\
1016  &  16.26576  &  2.19903  &  15.04  &  1.14  &  18.05  &  1.99  &  1.20  \\
NGC 6822  &    &    &    &    &   &    &     \\
12557  &  296.18396  &  -14.78028  &  17.12  &  1.11  &  19.22  &  0.56  &  2.14   \\
20331  &  296.20413  &  -14.63472  &  15.46  &  1.12  &  17.97  &  2.26  &  0.34   \\
10184  &  296.22358  &  -14.77472  &  15.13  &  1.31  &  17.97  &  1.53  &  1.75   \\
30133  &  296.21812  &  -14.88028  &  15.09  &  1.21  &  18.36  &  1.88  &  1.66   \\
20134  &  296.21321  &  -14.68083  &  15.07  &  1.28  &  18.12  &  2.76  &  0.35   \\
40139  &  296.20425  &  -14.74278  &  13.92  &  1.33  &  17.29  &  2.71  &  0.76   \\
10198  &  296.25087  &  -14.76778  &  14.19  &  1.31  &  17.08  &  2.08  &  1.09   \\
30292  &  296.18800  &  -14.87222  &  14.68  &  1.20  &  17.79  &  2.68  &  0.45   \\
10091  &  296.26700  &  -14.76306  &  14.18  &  1.28  &  17.23  &  2.46  &  0.73   \\
20004  &  296.21292  &  -14.73222  &  12.57  &  1.15  &  15.92  &  2.91  &  0.36   \\
12790  &  296.20389  &  -14.75535  &  16.59  &  1.44  &  18.91  &  1.30  &  1.62   \\
10817  &  296.23639  &  -14.83054  &  16.04  &  1.66  &  18.38  &  0.79  &  2.45   \\
20540  &  296.18011  &  -14.71067  &  16.39  &  1.49  &  19.01  &  1.58  &  1.59   \\
40590  &  296.28567  &  -14.73952  &  16.39  &  1.62  &  19.16  &  1.50  &  1.93   \\
12751  &  296.21027  &  -14.75973  &  16.29  &  1.60  &  18.79  &  1.17  &  2.08   \\
20578  &  296.29065  &  -14.69641  &  16.14  &  1.66  &  18.98  &  1.09  &  2.56   \\
20542  &  296.17703  &  -14.71031  &  16.02  &  1.82  &  19.28  &  1.88  &  2.07   \\
NGC 3109 &   &    &    &  &   &   &     \\
1067  &   150.73837  &  -26.14197  &  16.10  &  1.00  &  18.60  &  1.82  &  0.82   \\
1112  &   150.80582  &  -26.15483  &  16.04  &  1.14  &  18.24  &  1.44  &  1.11   \\
1153  &   150.75345  &  -26.17370  &  16.28  &  1.08  &  18.65  &  1.80  &  0.76   \\
2075  &   150.91414  &  -26.17125  &  16.56  &  1.61  &  19.31  &  1.46  &  1.96   \\
2081  &   150.86487  &  -26.15757  &  16.77  &  1.18  &  19.14  &  1.60  &  1.10   \\
3064  &   150.65652  &  -26.14415  &  15.84  &  1.15  &  18.61  &  1.93  &  1.05   \\
SMC  &    &    &    &    &   &   &      \\
IRAS00483--7347  &  12.52963  &  -73.52377  &  8.71  &  2.83  &  16.19  &  3.67  &  4.65   \\
GM103  &           12.62764  &  -72.85830  &  8.91  &  1.70  &  13.34  &  4.94  &  -0.82   \\
HV11417  &         15.20071  &  -72.85056  &  8.86  &  1.44  &  14.34  &  4.81  &  0.21   \\
LMC  &    &    &    &    &   &   &      \\
IRAS 04407--7000 & 070.118667 & -69.920417 &    8.69 &    1.96 &   15.45 &    5.84 &    0.51 \\
IRAS 04498--6842 & 072.422792 & -68.630944 &    7.49 &    1.64 &   13.04 &    5.68 &   -0.69 \\
IRAS 04509--6922 & 072.668583 & -69.292194 &    7.93 &    1.95 &   14.35 &    5.29 &    0.88 \\
IRAS 04516--6902 & 072.870792 & -68.963889 &    7.91 &    2.01 &   15.70 &    4.14 &    3.79 \\
MSX LMC 1210 & 073.835417 & -68.377583 &    9.72 &    2.94 &   16.44 &    3.25 &    4.51 \\
HV 888 & 076.058875 & -67.270639 &    6.78 &    1.23 &    9.53 &    2.66 &    0.14 \\
MSX LMC 642 & 082.200708 & -71.041361 &    9.88 &    3.12 &   14.74 &    5.18 &    0.27 \\
IRAS 05558--7000 & 088.837625 & -70.000833 &    9.13 &    2.86 &   17.60 &    4.41 &    4.69 \\

\end{tabular}
\begin{tablenotes}
   \small
   \item[*] Distance moduli used in Fig.~\ref{gaia_diag_sg}: IC 1613 (24.4), NGC 6822 (23.5), NGC 3109 (25.6), SMC (18.92), LMC (18.50).
   \end{tablenotes}
\end{threeparttable}
\label{tab_varGaia}
\end{table*}

\begin{table*}
        \caption{$JK_S$ and \textit{Gaia} ($BP$,$RP$) Photometry of supergiants and S stars in LG Galaxies}
\begin{threeparttable}
\centering
\begin{tabular}{cccccccc}
        Galaxy/Star \tnote{*}  & RA   & Dec   &  $K_S$  & $J-K_S$ & $RP$  & $BP-RP$   & $W_{RP}-W_K$        \\
         & \multicolumn{2}{c}{(2000.0)}   &  &  &  &  &  \\
IC 1613& supergiants &  &  &  &  &  & \\ 
1027 &  16.25295 &  2.18007 &  15.52  &  0.51  & 16.85 &   1.18 &   0.15 \\ 
4013 &  16.22231 &  2.08931 &  14.60  &  0.87  & 16.69 &   1.81 &   0.34 \\ 
1004 &  16.24328 &  2.15233 &  13.19  &  0.91  & 15.80 &   2.40 &   0.11 \\ 
1003 &  16.25696 &  2.14422 &  13.01  &  0.92  & 15.29 &   1.90 &   0.44 \\ 
1009 &  16.27144 &  2.19849 &  14.31  &  0.82  & 16.29 &   1.68 &   0.36 \\ 
1010 &  16.29004 &  2.20715 &  13.98  &  0.92  & 16.16 &   1.83 &   0.42 \\ 
1008 &  16.31963 &  2.18765 &  13.92  &  0.91  & 16.17 &   1.95 &   0.34 \\ 
3003 &  16.16047 &  2.01605 &  13.34  &  0.82  & 15.44 &   2.05 &  -0.01 \\ 
NGC 6822 &  supergiants&  &  &   &  &  &  \\ 
30016 &  296.19067 &  -14.87276 &  12.78  & 1.13  & 15.56 &   2.25 &   0.63 \\ 
40115 &  296.19919 &  -14.84817 &  13.26  & 1.15  & 16.08 &   2.37 &   0.52 \\ 
40177 &  296.21021 &  -14.73628 &  14.06  & 1.08  & 16.63 &   2.01 &   0.70 \\ 
10089 &  296.22278 &  -14.76476 &  14.02  & 1.08  & 16.50 &   1.88 &   0.78 \\ 
10032 &  296.22696 &  -14.80191 &  13.34  & 1.14  & 16.07 &   2.25 &   0.59 \\ 
40278 &  296.22726 &  -14.85778 &  12.33  & 1.10  & 15.08 &   2.27 &   0.55 \\ 
40315 &  296.23212 &  -14.86564 &  12.33  & 1.06  & 15.11 &   2.38 &   0.41 \\ 
10011 &  296.23883 &  -14.82247 &  12.46  & 1.14  & 15.29 &   2.26 &   0.67 \\ 
10015 &  296.24945 &  -14.75443 &  12.70  & 0.94  & 14.99 &   1.96 &   0.39 \\ 
20101 &  296.26492 &  -14.72723 &  14.81  & 1.15  & 17.38 &   1.88 &   0.91 \\ 
NGC 6822 & S stars&  &  &   &  &  &  \\ 
10870 &  296.17892 &  -14.82286 &  16.19  & 1.34  & 18.98 &   1.44 &   1.84 \\ 
10784 &  296.21545 &  -14.83469 &  16.20  & 1.26  & 18.53 &   1.24 &   1.58 \\ 
11004 &  296.27341 &  -14.80861 &  16.27  & 1.33  & 18.77 &   1.12 &   1.97 \\ 
11029 &  296.28308 &  -14.80497 &  16.22  & 1.24  & 18.66 &   1.00 &   1.99 \\ 
30528 &  296.19156 &  -14.89296 &  16.51  & 1.26  & 19.27 &   1.60 &   1.55 \\ 
10326 &  296.25522 &  -14.82579 &  15.52  & 1.30  & 18.21 &   1.57 &   1.54 \\ 

\end{tabular}
\begin{tablenotes}
\small
\item[*] Distance moduli used in Fig.~\ref{gaia_diag_sg}: IC 1613 (24.4), NGC 6822 (23.5).
\end{tablenotes}
\end{threeparttable}
\label{tab_sgGaia}
\end{table*}

\begin{figure*}
	\includegraphics[width=0.80\textwidth]{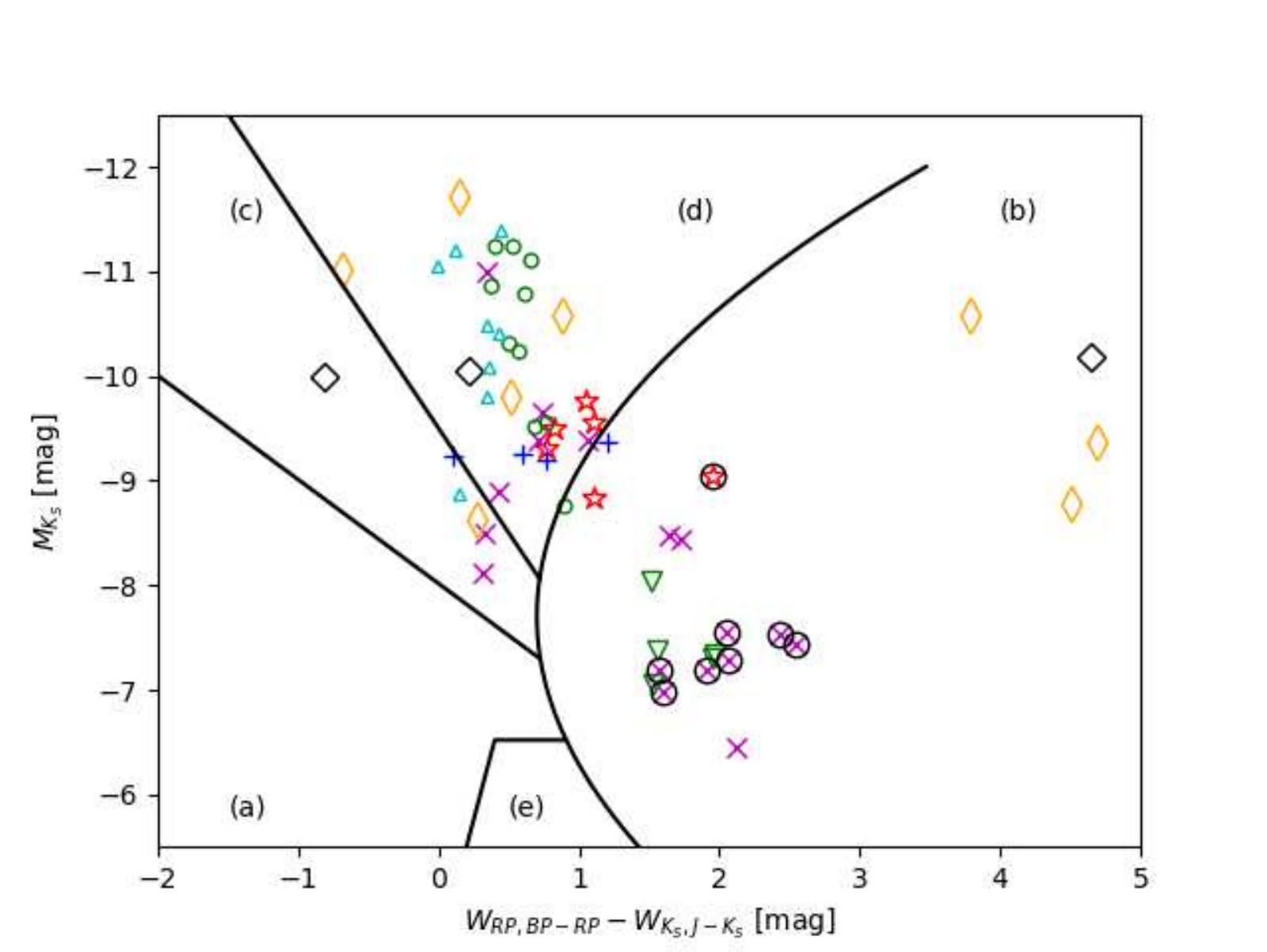}
	\caption{Modified \citet{Lebzelter_2018} diagram for long-period variables in NGC~3109 ($\ast$), IC~1613 (+) including supergiants ($\bigtriangleup$), NGC~6822 (x) including supergiants ($\circ$) and S stars ($\bigtriangledown$), three very long period variables in the SMC ($\diamond$) and eight very long period O-rich variables in the LMC (thin $\diamond$). A large circle is placed around the symbols for the known or suspected C stars for emphasis. The annotated regions indicate where stars of a particular type are found, viz., (a) low mass O-rich AGB stars, (b) C-rich stars, (c) O-rich AGB stars of intermediate mass, (d) red supergiants and O-rich massive AGB stars, and (e) RGB and faint AGB stars. Absolute $K_S$ magnitude is plotted on the y axis using $(m-M)_0 = 18.5$ for the LMC and published values for the other galaxies.} 
\label{gaia_diag_sg}
\end{figure*}

	The data for the Miras are tabulated in Table~\ref{tab_varGaia} where $BP$ and $RP$ are \textit{Gaia} DR2 blue and red magnitudes, and $W_{RP}-W_K$ are 'colours' as defined in \citet{Lebzelter_2018}. The supergiants and S stars are in Table~\ref{tab_sgGaia}. In Fig.~\ref{gaia_diag_sg} we plot M$_{K_S}$ against $W_{RP}-W_K$. The solid lines separate the regions defined by \citet{Lebzelter_2018}; they are to some extent arbitrary and points lying close to them  cannot be confidently assigned to a particular region. We have included known supergiants in the galaxies in the diagram. Absolute $K_S$ magnitude is plotted on the y axis in contract to apparent LMC $K_S$ magnitude in \citet{Lebzelter_2018}.

	Because the stars are relatively faint, \textit{Gaia} photometry is not available for them all. In IC 1613 only the four brightest O-rich Miras were found. These were considered by \citet{Menzies_2015} to be massive stars undergoing HBB, and they appear in Fig.~\ref{gaia_diag_sg} in region (b) as expected. All the unambiguous red supergiants cluster in region (d), while the Cepheid, v25, lies in region (c).

For NGC 6822 there are data for seven of the 50 C-rich Miras \citep{Whitelock_2013} which all fall in region (b), and 10 of the 11 O-rich Miras. The four HBB stars and the possible red supergiant appear in region (d), and two of the other O-rich variables are in region (c). However, the remaining three are unexpectedly in region (b) amongst the C-rich stars. As discussed by \citet{Lebzelter_2018} such stars might have dust shells which contribute extra reddening not accounted for in the Wesenheit formulae. The non-variable red supergiants appear in region (d), while the S stars are to be found in the lower C star region as \citet{Lebzelter_2018} suggested they would be.

Six of the eight Miras in NGC 3109 reported in this paper have data. Four of the suspected HBB stars fall in region (d) as would be expected, while a fifth is in region (b), close to the boundary with region (d). The very long period star (\#2075) lies in region (b) and may be C-rich as speculated in section~\ref{sec_evol}.

The SMC stars are all O-rich. Two appear in regions (a) and (d), while the third is in region (b) where extreme C stars are expected. In the last case, there must be a very thick dust shell providing extra reddening.

For the LMC stars, we only have single-epoch 2MASS measurements. These stars are likely to have relatively large amplitudes so the positions in Fig.~\ref{gaia_diag_sg} in the vertical direction are somewhat more uncertain than for the stars in the other galaxies. Nevertheless, three of the LMC stars lie in a similar position to the single extremely red SMC Mira and, like it, probably have thick dust shells.

None of the WLM variables have \textit{Gaia} DR2 data so could not be included.

Thus, most of this small sample of Miras in Local Group galaxies appear in the Lebzelter diagram where predicted on the basis of the LMC data. However, this diagram can only be used as supporting evidence for the chemical characteristics of a specific variable; spectra are necessary for a definite attribution to be made.

\section{Very Long Period Variables in the SMC}\label{sec_LPV_SMC}
The measurements reported here were obtained with the 1.9-m telescope at SAAO and are on the SAAO photometric system as defined by \citet{Carter1990}, i.e. they are different from the IRSF photometry reported in the body of the paper (\citet{Carpenter_2001} provides transformation equations between the SAAO and 2MASS systems). The errors are less than 0.03 mag at $JHK$ and less than 0.05 mag at $L$ except where marked with a colon where they are less than 0.1 mag. Fourier mean values for the photometry are given in Table~\ref{tab_SMC}.

\begin{table}
\centering
\caption{HV11417}
\begin{tabular}{crrrl}
	\hline
 JD & $J$ & $H$ & $K$ & $L$\\
$-2440000$ & \multicolumn{4}{c}{(mag)}\\
\hline
0391.38& 10.37&  9.44&  8.93 & 8.41\\
0446.31& 10.57 & 9.59 & 9.06 & 8.51:\\
0499.28& 10.75 & 9.76 & 9.22 & 8.60:\\
0702.60& 11.07& 10.11 & 9.49 & 8.64:\\
0761.30& 11.07& 10.15 & 9.52 & 8.79:\\
0795.31& 10.65&  9.82 & 9.30 & 8.61:\\
1012.72&   9.86&  8.87 & 8.52 & 7.98\\
1210.33&   9.65&  8.65&  8.27 & 7.84\\
1451.50& 10.14&  9.91  &8.71 & 8.25\\
1481.39&  10.27&  9.27 & 8.80 & 8.30\\
1539.29& 10.43 & 9.41 & 8.90 & 8.45:\\
\hline
\end{tabular}
\end{table}

\begin{table}
\centering
\caption{IRAS00483--7347}
\begin{tabular}{crrrl}
	\hline
 JD & $J$ & $H$ & $K$ & $L$\\
$-2440000$ & \multicolumn{4}{c}{(mag)}\\
\hline
6658.56&	11.05  &	9.45	    &   8.44&	7.18\\
6686.47&	11.09&	9.46&	8.45	    &    7.19\\
6704.49&	11.02&	9.47&	8.45&	\\
6721.36&	11.08&	9.48	   &    8.46&	7.21\\
6746.46&	11.11&	9.53&	8.53&	7.29\\
6783.29&	11.19&   9.52&	8.49&	7.27\\
6841.27&	11.17&	9.53&	8.50&	7.23\\
6981.64&	11.48	&9.83&	8.75&	7.46\\
7078.43&	11.62&	9.94&	8.85&	7.53\\
7096.33&	11.69	&9.97&	8.85&	\\
7120.32&	11.69&	9.97&	8.85&	7.54\\
7133.31&	11.71&	10.11&	8.88&	7.56\\
7160.31&	11.76&	10.05&	8.92&	\\
7362.53&	12.14&	10.28&	9.04&	7.55\\
7383.63&	12.12 &	10.28&	9.02&	7.58\\
7425.45&	11.98&	10.16&	8.93&	7.50\\
7462.47&	11.92&	10.07&	8.82&	7.38\\
7484.41&	11.82&	10.01&	8.80&	7.31\\
7767.48&	11.79&	10.03&	8.85&	7.42\\
7782.41&	         &	         &	8.99&	\\
7838.44&	11.85&	10.13&	8.95&	7.47\\
8592.32&	11.28&	9.46&	8.38&	6.92\\
\hline
\end{tabular}
\end{table}

\begin{table}
\centering
\caption{GM103}
\begin{tabular}{crrrl}
	\hline
 JD & $J$ & $H$ & $K$ & $L$\\
$-2440000$ & \multicolumn{4}{c}{(mag)}\\
\hline
387.32&	11.25&	10.08&	9.39&	8.62\\
442.38&	10.95&	9.89&	9.21&	8.34\\
500.27&	10.45&	9.47&	8.86&	7.88\\
705.49&	10.05&	8.91&	8.38&	7.50\\
762.3&    10.02&	8.87	  &     8.34&	7.53\\
793.3&	9.99&	8.88&	8.34&	7.50\\
1013.72&	10.33&	9.26&	8.70&	8.01\\
1181.72&	10.90&	9.90	  &     9.24&	8.49\\
1209.32&	11.09&	10.00&	9.35&	8.72\\
1391.66&	11.22&	10.11&	9.44&	8.71\\
1453.52&	11.26&	10.14&	9.43&	8.67\\
1482.33&	11.21&	10.13&	9.43&	8.59\\
1538.29&	10.56&	9.65	&     8.98&	8.12\\
1712.69&	10.28&	9.14&	8.57	  &     7.73\\
1813.46	&10.19&	8.97&	8.40&	7.61\\
\hline
\end{tabular}
\end{table}

\section{Photometric data for NGC~3109}\label{sec_Phot}
The time-series photometry for the variables  reported on in this paper is presented in the following nine tables, Tables~\ref{tab_1060 } to \ref{tab_3089 }. Column contents are indicated by the headings and are obvious except that $\epsilon_J$, etc are photometric errors.
A sample of the photometric data catalogue is shown in the accompanying Table~\ref{tab_CMall}.
The full photometric catalogue  and the variable star files can be obtained online.

\begin{table}
 \caption{Photometric data for variable \#1060 }
 \centering
 \begin{tabular}{lcccccc}
 JD & $J$ & $\epsilon_J$ & $H$ & $\epsilon_H$ & $K_S$ & $\epsilon_K$ \\
2452265.49565 &   15.675 & 0.041 &  15.722 & 0.010 &  15.917 & 0.008 \\
2452349.50323 &   15.557 & 0.022 &  15.691 & 0.011 &  15.878 & 0.009 \\
2452412.50586 &   15.553 & 0.015 &  15.560 & 0.014 &  15.778 & 0.010 \\
2452671.54266 &   15.409 & 0.023 &  15.443 & 0.013 &  15.704 & 0.013 \\
2452704.31954 &   15.373 & 0.017 &  15.445 & 0.008 &  15.641 & 0.008 \\
2452707.27169 &   15.437 & 0.027 &  15.389 & 0.010 &  15.627 & 0.008 \\
2452709.50897 &   15.370 & 0.014 &  15.429 & 0.009 &  15.649 & 0.008 \\
2452710.52247 &   15.351 & 0.018 &  15.430 & 0.009 &  15.665 & 0.008 \\
2452755.38687 &   15.331 & 0.011 &  15.402 & 0.010 &  15.645 & 0.009 \\
2452783.35084 &   15.326 & 0.011 &  15.407 & 0.009 &  15.645 & 0.007 \\
2452809.26936 &   15.293 & 0.015 &  15.426 & 0.010 &  15.646 & 0.007 \\
2452967.58335 &   15.225 & 0.017 &  15.272 & 0.010 &  15.555 & 0.009 \\
2453013.43395 &   15.182 & 0.017 &  15.268 & 0.008 &  15.530 & 0.006 \\
2453173.26920 &   15.116 & 0.010 &  15.258 & 0.008 &  15.507 & 0.009 \\
2453350.52446 &   15.031 & 0.014 &  15.180 & 0.007 &  15.417 & 0.005 \\
2453438.42628 &   15.033 & 0.013 &  15.159 & 0.008 &  15.418 & 0.010 \\
\end{tabular} 
\label{tab_1060 }
 \end{table}

\begin{table}
 \caption{Photometric data for variable \#1067 }
 \centering
 \begin{tabular}{lcccccc}
 JD & $J$ & $\epsilon_J$ & $H$ & $\epsilon_H$ & $K_S$ & $\epsilon_K$ \\
2452265.49565 &   15.755 & 0.041 &  16.028 & 0.014 &  16.864 & 0.017 \\
2452349.50323 &   16.149 & 0.031 &  16.560 & 0.019 &  17.239 & 0.020 \\
2452412.50586 &   16.394 & 0.029 &  16.830 & 0.017 &  17.456 & 0.018 \\
2452671.54266 &   16.165 & 0.052 &  16.555 & 0.025 &  17.105 & 0.034 \\
2452704.31954 &   16.240 & 0.039 &  16.417 & 0.017 &  17.130 & 0.029 \\
2452707.27169 &   16.150 & 0.043 &  16.336 & 0.026 &  17.056 & 0.025 \\
2452709.50897 &   16.116 & 0.027 &  16.369 & 0.014 &  17.046 & 0.016 \\
2452710.52247 &   16.055 & 0.028 &  16.377 & 0.014 &  17.027 & 0.016 \\
2452755.38687 &   15.886 & 0.014 &  16.185 & 0.010 &  16.967 & 0.011 \\
2452783.35084 &   15.829 & 0.015 &  16.157 & 0.012 &  16.911 & 0.009 \\
2452809.26936 &   15.784 & 0.018 &  16.165 & 0.014 &  16.889 & 0.014 \\
2452967.58335 &   16.338 & 0.043 &  16.756 & 0.027 &  17.372 & 0.026 \\
2453013.43395 &   16.777 & 0.078 &  17.178 & 0.041 &  17.584 & 0.032 \\
2453173.26920 &   16.506 & 0.035 &  16.914 & 0.029 &  17.585 & 0.028 \\
2453350.52446 &   15.864 & 0.028 &  16.010 & 0.012 &  16.740 & 0.014 \\
2453438.42628 &   15.626 & 0.017 &  15.925 & 0.009 &  16.709 & 0.014 \\
\end{tabular} 
\label{tab_1067 }
 \end{table}
 
\begin{table}
 \caption{Photometric data for variable \#1112 }
 \centering
 \begin{tabular}{lcccccc}
 JD & $J$ & $\epsilon_J$ & $H$ & $\epsilon_H$ & $K_S$ & $\epsilon_K$ \\
2452265.49565 &   16.061 & 0.061 &  16.380 & 0.028 &  17.181 & 0.028 \\
2452349.50323 &   16.193 & 0.037 &  16.494 & 0.029 &  17.276 & 0.028 \\
2452412.50586 &   16.314 & 0.037 &  16.567 & 0.033 &  17.396 & 0.029 \\
2452671.54266 &   15.713 & 0.035 &  16.072 & 0.025 &  16.911 & 0.032 \\
2452704.31954 &   15.974 & 0.035 &  16.310 & 0.021 &  17.111 & 0.029 \\
2452707.27169 &   15.940 & 0.039 &  16.227 & 0.026 &  17.018 & 0.025 \\
2452709.50897 &   15.941 & 0.028 &  16.281 & 0.023 &  17.034 & 0.023 \\
2452710.52247 &   15.874 & 0.028 &  16.281 & 0.023 &  17.055 & 0.024 \\
2452755.38687 &   16.062 & 0.028 &  16.474 & 0.028 &  17.216 & 0.024 \\
2452783.35084 &   16.097 & 0.026 &  16.568 & 0.028 &  17.269 & 0.023 \\
2452809.26936 &   16.176 & 0.034 &  16.573 & 0.029 &  17.309 & 0.028 \\
2452967.58335 &   16.138 & 0.041 &  16.497 & 0.029 &  17.261 & 0.030 \\
2453013.43395 &   16.357 & 0.068 &  16.701 & 0.037 &  17.539 & 0.041 \\
2453173.26920 &   15.589 & 0.021 &  15.932 & 0.019 &  16.800 & 0.019 \\
2453350.52446 &   16.054 & 0.045 &  16.402 & 0.024 &  17.145 & 0.025 \\
2453438.42628 &   16.133 & 0.039 &  16.540 & 0.032 &  17.300 & 0.031 \\
\end{tabular} 
\label{tab_1112 }
 \end{table}
 
\begin{table}
 \caption{Photometric data for variable \#1153 }
 \centering
 \begin{tabular}{lcccccc}
 JD & $J$ & $\epsilon_J$ & $H$ & $\epsilon_H$ & $K_S$ & $\epsilon_K$ \\
  2265.49565 &   15.895 & 0.045 &  16.163 & 0.018 &  16.900 & 0.018 \\
  2349.50323 &   15.662 & 0.022 &  16.146 & 0.015 &  16.845 & 0.017 \\
  2412.50586 &   15.816 & 0.021 &  16.200 & 0.015 &  16.958 & 0.014 \\
  2671.54266 &   16.482 & 0.068 &  16.807 & 0.035 &  17.853 & 0.074 \\
  2704.31954 &   16.425 & 0.048 &  16.939 & 0.032 &  17.679 & 0.052 \\
  2707.27169 &   16.538 & 0.056 &  16.917 & 0.041 &  17.541 & 0.035 \\
  2709.50897 &   16.473 & 0.039 &  16.867 & 0.020 &  17.556 & 0.028 \\
  2710.52247 &   16.467 & 0.048 &  16.950 & 0.025 &  17.633 & 0.028 \\
  2755.38687 &   16.585 & 0.028 &  16.957 & 0.018 &  17.666 & 0.020 \\
  2783.35084 &   16.461 & 0.023 &  16.919 & 0.024 &  17.522 & 0.019 \\
  2809.26936 &   16.435 & 0.037 &  16.793 & 0.028 &  17.430 & 0.026 \\
  2967.58335 &   15.771 & 0.027 &  16.026 & 0.017 &  16.756 & 0.017 \\
  3013.43395 &   15.743 & 0.034 &  16.057 & 0.014 &  16.807 & 0.018 \\
  3173.26920 &   16.310 & 0.028 &  16.891 & 0.028 &  17.354 & 0.026 \\
  3350.52446 &   16.776 & 0.068 &  17.047 & 0.032 &  17.567 & 0.029 \\
  3438.42628 &   16.705 & 0.041 &  17.225 & 0.032 &  17.803 & 0.032 \\
\end{tabular} 
\label{tab_1153 }
 \end{table}

\begin{table}
 \caption{Photometric data for variable \#1224 }
 \centering
 \begin{tabular}{lcccccc}
 JD & $J$ & $\epsilon_J$ & $H$ & $\epsilon_H$ & $K_S$ & $\epsilon_K$ \\
2452265.49565 &   16.534 & 0.082 &  17.082 & 0.041 &  17.666 & 0.034 \\
2452349.50323 &   16.381 & 0.039 &  16.770 & 0.022 &  17.378 & 0.023 \\
2452412.50586 &   16.464 & 0.029 &  16.820 & 0.015 &  17.492 & 0.018 \\
2452671.54266 &   15.848 & 0.039 &  16.290 & 0.025 &  17.047 & 0.032 \\
2452704.31954 &   16.090 & 0.032 &  16.447 & 0.018 &  17.172 & 0.028 \\
2452707.27169 &   15.940 & 0.037 &  16.371 & 0.026 &  17.136 & 0.024 \\
2452709.50897 &   16.043 & 0.028 &  16.460 & 0.015 &  17.140 & 0.016 \\
2452710.52247 &   16.055 & 0.030 &  16.418 & 0.017 &  17.143 & 0.018 \\
2452755.38687 &   16.247 & 0.024 &  16.706 & 0.015 &  17.324 & 0.015 \\
2452783.35084 &   16.374 & 0.025 &  16.849 & 0.024 &  17.476 & 0.021 \\
2452809.26936 &   16.528 & 0.035 &  17.068 & 0.035 &  17.646 & 0.030 \\
2452967.58335 &   16.584 & 0.054 &  17.252 & 0.043 &  17.875 & 0.041 \\
2453013.43395 &   16.651 & 0.070 &  17.063 & 0.039 &  17.754 & 0.039 \\
2453173.26920 &   16.265 & 0.028 &  16.725 & 0.027 &  17.313 & 0.024 \\
2453350.52446 &   15.988 & 0.032 &  16.409 & 0.018 &  17.049 & 0.018 \\
2453438.42628 &   16.311 & 0.032 &  16.861 & 0.025 &  17.496 & 0.028 \\
\end{tabular} 
\label{tab_1224 }
 \end{table}

\begin{table}
 \caption{Photometric data for variable \#2075 }
 \centering
 \begin{tabular}{lcccccc}
 JD & $J$ & $\epsilon_J$ & $H$ & $\epsilon_H$ & $K_S$ & $\epsilon_K$ \\
2452265.49564 &   15.847 & 0.025 &  16.395 & 0.024 &  17.255 & 0.028 \\
2452349.50322 &     --   &  --   &  16.548 & 0.018 &  17.619 & 0.026 \\
2452412.50586 &   16.192 & 0.025 &  16.671 & 0.014 &  17.843 & 0.018 \\
2452671.54854 &   16.948 & 0.093 &  17.373 & 0.052 &  18.706 & 0.109 \\
2452704.33801 &   17.204 & 0.103 &  17.393 & 0.043 &  18.749 & 0.119 \\
2452707.30243 &   17.019 & 0.097 &  17.449 & 0.046 &  18.825 & 0.092 \\
2452709.53763 &   16.963 & 0.061 &  17.566 & 0.039 &  18.932 & 0.081 \\
2452710.55110 &   16.951 & 0.075 &  17.511 & 0.046 &  18.943 & 0.089 \\
2452755.42294 &   17.014 & 0.039 &  17.629 & 0.021 &  19.004 & 0.043 \\
2452783.37426 &   17.140 & 0.054 &  17.751 & 0.048 &  19.198 & 0.081 \\
2452809.29359 &   17.298 & 0.089 &  17.707 & 0.076 &  18.921 & 0.097 \\
2452968.55510 &   16.459 & 0.041 &  16.771 & 0.024 &  17.674 & 0.034 \\
2453013.46903 &   16.097 & 0.043 &  16.565 & 0.019 &  17.450 & 0.025 \\
2453173.29094 &   15.715 & 0.018 &  16.207 & 0.017 &  17.185 & 0.017 \\
2453350.54581 &   15.894 & 0.029 &  16.316 & 0.014 &  17.236 & 0.018 \\
2453438.44598 &   15.991 & 0.019 &  16.478 & 0.014 &  17.521 & 0.017 \\
2453533.26144 &   16.158 & 0.028 &  16.603 & 0.019 &  17.741 & 0.027 \\
\end{tabular} 
\label{tab_2075 }
 \end{table}

\begin{table}
 \caption{Photometric data for variable \#2081 }
 \centering
 \begin{tabular}{lcccccc}
 JD & $J$ & $\epsilon_J$ & $H$ & $\epsilon_H$ & $K_S$ & $\epsilon_K$ \\
2452265.49564 &   16.987 & 0.064 &  17.174 & 0.041 &  18.109 & 0.061 \\
2452349.50322 &     --   &  --   &  17.334 & 0.035 &  18.189 & 0.046 \\
2452412.50586 &   17.075 & 0.051 &  17.295 & 0.022 &  17.963 & 0.021 \\
2452671.54854 &   16.573 & 0.071 &  16.996 & 0.037 &  17.877 & 0.057 \\
2452704.33801 &   16.820 & 0.072 &  17.377 & 0.043 &  18.369 & 0.093 \\
2452707.30243 &   16.819 & 0.075 &  17.210 & 0.037 &  18.174 & 0.046 \\
2452709.53763 &   16.700 & 0.048 &  17.193 & 0.028 &  18.075 & 0.034 \\
2452710.55110 &   16.834 & 0.057 &  17.268 & 0.029 &  18.001 & 0.037 \\
2452755.42294 &   17.008 & 0.043 &  17.378 & 0.019 &  18.233 & 0.025 \\
2452783.37426 &   16.994 & 0.043 &  17.283 & 0.032 &  18.043 & 0.028 \\
2452809.29359 &   16.867 & 0.056 &  17.272 & 0.052 &  17.987 & 0.037 \\
2452968.55510 &   16.352 & 0.039 &  16.600 & 0.019 &  17.397 & 0.024 \\
2453013.46903 &   16.421 & 0.054 &  16.688 & 0.021 &  17.498 & 0.028 \\
2453173.29094 &   16.957 & 0.056 &  17.312 & 0.043 &  18.201 & 0.046 \\
2453350.54581 &   16.675 & 0.052 &  16.991 & 0.026 &  17.690 & 0.030 \\
2453438.44598 &   16.454 & 0.032 &  16.665 & 0.018 &  17.496 & 0.018 \\
2453533.26144 &   16.721 & 0.045 &  17.049 & 0.025 &  17.863 & 0.029 \\
\end{tabular} 
\label{tab_2081 }
 \end{table}
 
\begin{table}
 \caption{Photometric data for variable \#3064 }
 \centering
 \begin{tabular}{lcccccc}
 JD & $J$ & $\epsilon_J$ & $H$ & $\epsilon_H$ & $K_S$ & $\epsilon_K$ \\
2452265.49731 &   15.897 & 0.043 &  16.260 & 0.019 &  17.152 & 0.023 \\
2452349.50466 &   15.685 & 0.028 &  15.863 & 0.012 &  16.785 & 0.017 \\
2452412.50517 &   15.526 & 0.015 &  15.837 & 0.009 &  16.721 & 0.009 \\
2452671.55656 &   15.898 & 0.034 &  16.211 & 0.017 &  16.989 & 0.027 \\
2452704.35866 &   16.010 & 0.035 &  16.213 & 0.018 &  17.105 & 0.032 \\
2452707.32826 &   15.943 & 0.027 &  16.234 & 0.015 &  17.012 & 0.014 \\
2452709.56404 &   15.921 & 0.028 &  16.193 & 0.016 &  17.043 & 0.018 \\
2452755.44707 &   16.066 & 0.018 &  16.292 & 0.010 &  17.179 & 0.010 \\
2452809.30990 &   15.985 & 0.026 &  16.266 & 0.021 &  17.131 & 0.019 \\
2452968.58457 &   15.561 & 0.015 &  15.880 & 0.009 &  16.787 & 0.013 \\
2453013.50454 &   15.622 & 0.023 &  15.921 & 0.009 &  16.838 & 0.014 \\
2453093.27710 &   15.939 & 0.037 &  16.281 & 0.018 &  17.258 & 0.026 \\
2453173.30900 &   15.974 & 0.023 &  16.210 & 0.020 &  17.066 & 0.018 \\
2453350.56978 &   16.109 & 0.029 &  16.364 & 0.014 &  17.196 & 0.017 \\
2453438.47422 &   15.800 & 0.018 &  16.004 & 0.010 &  16.850 & 0.015 \\
2453533.27828 &   15.613 & 0.016 &  15.951 & 0.012 &  16.797 & 0.014 \\
\end{tabular} 
\label{tab_3064 }
 \end{table}

\begin{table}
 \caption{Photometric data for variable \#3089 }
 \centering
 \begin{tabular}{lcccccc}
 JD & $J$ & $\epsilon_J$ & $H$ & $\epsilon_H$ & $K_S$ & $\epsilon_K$ \\
2452265.49731 &   16.603 & 0.086 &  17.175 & 0.048 &  18.006 & 0.050 \\
2452349.50466 &   16.469 & 0.050 &  17.175 & 0.041 &  18.112 & 0.054 \\
2452412.50517 &   16.615 & 0.035 &  17.176 & 0.018 &  18.038 & 0.023 \\
2452671.55656 &   15.722 & 0.029 &  16.291 & 0.019 &  17.174 & 0.030 \\
2452704.35866 &   15.762 & 0.028 &  16.249 & 0.018 &  17.163 & 0.035 \\
2452707.32826 &   15.767 & 0.024 &  16.229 & 0.014 &  17.091 & 0.016 \\
2452709.56404 &   15.687 & 0.024 &  16.130 & 0.013 &  17.058 & 0.018 \\
2452755.44707 &   15.590 & 0.014 &  16.124 & 0.009 &  17.064 & 0.010 \\
2452809.30990 &   15.534 & 0.018 &  16.020 & 0.017 &  16.913 & 0.017 \\
2452968.58457 &   15.609 & 0.020 &  16.042 & 0.010 &  16.946 & 0.014 \\
2453013.50454 &   15.641 & 0.025 &  16.056 & 0.010 &  16.989 & 0.014 \\
2453093.27710 &   15.605 & 0.028 &  16.141 & 0.014 &  17.152 & 0.023 \\
2453173.30900 &   15.689 & 0.018 &  16.270 & 0.018 &  17.233 & 0.021 \\
2453350.56978 &   15.981 & 0.027 &  16.595 & 0.017 &  17.600 & 0.021 \\
2453438.47422 &   16.122 & 0.024 &  16.771 & 0.020 &  17.773 & 0.026 \\
2453533.27828 &   16.300 & 0.029 &  16.913 & 0.026 &  17.990 & 0.034 \\
\end{tabular} 
\label{tab_3089 }
 \end{table}

\begin{table}
\caption{Photometric Data for NGC~3109}
\begin{threeparttable}
\centering
\begin{tabular}{ccccccccccccc}

	RA & Dec & N \tnote{a}& $J$ & $\sigma_J$\tnote{b} & $H$ & $\sigma_H$\tnote{b} & $K_S$ & $\sigma_K$ \tnote{b}& $J-H$ & $H-K_S$ & $J-K_S $&Note \tnote{c} \\ 
	\multicolumn{2}{c}{(J2000.0)} & &\multicolumn{10}{c}{}\\
150.75345 & -26.17370 & 1153 & 17.367 & 0.379 & 16.682 & 0.405 & 16.284 & 0.371 & 0.685 & 0.398 & 1.083 & V   \\
150.80502 & -26.16340 & 1171 & 18.258 & 0.062 & 17.503 & 0.104 & 17.463 & 0.284 & 0.755 & 0.040 & 0.795 &    \\
150.81224 & -26.16286 & 1173 & 18.018 & 0.110 & 17.109 & 0.079 & 16.704 & 0.125 & 0.909 & 0.405 & 1.314 & C   \\
150.79617 & -26.16277 & 1174 & 17.823 & 0.061 & 17.080 & 0.046 & 16.973 & 0.116 & 0.743 & 0.107 & 0.850 &    \\
150.80377 & -26.16170 & 1177 & 18.196 & 0.145 & 17.310 & 0.123 & 16.752 & 0.110 & 0.886 & 0.558 & 1.444 & G   \\
\end{tabular}
\begin{tablenotes}
\small
\item[a] catalogue star number
\item[b] standard deviation of tabulated magnitude	
\item[c] V indicates \textit{variable}; C indicates \textit{C star}; G indicates \textit{probable background galaxy}
\end{tablenotes}
\end{threeparttable}
\label{tab_CMall}
\end{table}

\end{document}